\documentclass[twocolumn]{pasj01}

\usepackage[switch,mathlines]{lineno}

\Received{2023/12/20}
\Accepted{2024/01/26}

\usepackage{natbib}
\bibpunct[:]{(}{)}{;}{a}{}{,}

\usepackage{xcolor}
\usepackage{mathpazo}
\usepackage[T1]{fontenc} 
\usepackage{appendix}

\begin{document}

\title{Optical and soft X-ray light-curve analysis during the 2022 eruption of U Scorpii:\\ structural changes in the accretion disk}

\author{
Katsuki \textsc{Muraoka},\altaffilmark{1,}\footnotemark[*] 
Naoto \textsc{Kojiguchi},\altaffilmark{1}
Junpei \textsc{Ito},\altaffilmark{1}
Daisaku \textsc{Nogami},\altaffilmark{1} 
Taichi \textsc{Kato},\altaffilmark{1}
Yusuke \textsc{Tampo},\altaffilmark{1} 
Kenta \textsc{Taguchi},\altaffilmark{1}
Keisuke \textsc{Isogai},\altaffilmark{1,2,3}
Teofilo \textsc{Arranz},\altaffilmark{4}
John \textsc{Blackwell},\altaffilmark{5}             
David \textsc{Blane},\altaffilmark{6,7} \\
Stephen \textsc{M. Brincat},\altaffilmark{8}
Graeme \textsc{Coates},\altaffilmark{9}            
Walter \textsc{Cooney},\altaffilmark{10} \\           
Shawn \textsc{Dvorak},\altaffilmark{11}                 
Charles \textsc{Galdies},\altaffilmark{12}          
Daniel \textsc{Glomski},\altaffilmark{13} \\
Franz-Josef \textsc{Hambsch},\altaffilmark{14,15,16}    %
Barbara \textsc{Harris},\altaffilmark{17}               
John \textsc{Hodge},\altaffilmark{18} \\
Jose L. \textsc{Hernández-Verdejo},\altaffilmark{19}       
Marco \textsc{Iozzi},\altaffilmark{20}              
Hiroshi \textsc{Itoh},\altaffilmark{21} \\              %
Seiichiro \textsc{Kiyota},\altaffilmark{22} 
Darrell \textsc{Lee},\altaffilmark{23} 
Magnus \textsc{Larsson},\altaffilmark{24}           
Tapio \textsc{Lahtinen},\altaffilmark{25}           
Gordon \textsc{Myers},\altaffilmark{26}                 
Berto \textsc{Monard},\altaffilmark{27,28}
Mario \textsc{Morales Aimar},\altaffilmark{29} \\   
Masayuki \textsc{Moriyama},\altaffilmark{30}            %
Masanori \textsc{Mizutani},\altaffilmark{31}            %
Kazuo \textsc{Nagai},\altaffilmark{32} \\               %
Thabet \textsc{AlQaissieh},\altaffilmark{33} 
Aldrin \textsc{B.Gabuya},\altaffilmark{33}
Mohammad \textsc{Odeh},\altaffilmark{34} \\
Carlos \textsc{Perello},\altaffilmark{35}
Andrew \textsc{Pearce},\altaffilmark{36} 
Juan \textsc{Miguel Perales},\altaffilmark{35,37} \\
David \textsc{Quiles},\altaffilmark{38}                
Filipp \textsc{D. Romanov},\altaffilmark{39,40,41,42,43}
David \textsc{J. Lane},\altaffilmark{40,41,42}
Michael \textsc{Richmond},\altaffilmark{44}         
Nello \textsc{Ruocco},\altaffilmark{45} 
Yasuo \textsc{Sano},\altaffilmark{46,47,48} \\
Mark \textsc{Spearman},\altaffilmark{49,50}
Richard \textsc{Schmidt},\altaffilmark{51}    
Tonny \textsc{Vanmunster},\altaffilmark{16,52,53} 
Pavol \textsc{A. Dubovsky},\altaffilmark{54} 
Richard \textsc{Wagner},\altaffilmark{55,56}        
Guido \textsc{Wollenhaupt},\altaffilmark{57,58}            
Joachim \textsc{Lorenz},\altaffilmark{58} 
Gerhard \textsc{Lehmann},\altaffilmark{58} 
Andrea \textsc{Salewski},\altaffilmark{58} and \\
Guy \textsc{Williamson}\altaffilmark{59}            
}

\altaffiltext{1}{Department of Astronomy, Graduate School of Science, Kyoto University, Kitashirakawa-Oiwake-cho, Sakyo-ku, Kyoto-shi, Kyoto 606-8502, Japan}
\altaffiltext{2}{Okayama Observatory, Kyoto University, 3037-5 Honjo, Kamogata-cho, Asakuchi-shi, Okayama 719-0232, Japan}
\altaffiltext{3}{Department of Multi-Disciplinary Sciences, Graduate School of Arts and Sciences, The University of Tokyo, 3-8-1 Komaba, Meguro-ku, Tokyo 153-8902, Japan}
\altaffiltext{4}{Observatorio Astronómico "Las Pegueras", Navas De Oro, Segovia, Spain}
\altaffiltext{5}{Phillips Exeter Academy, 20 Main Street, Exeter, New Hampshire 03833, USA}
\altaffiltext{6}{American Association of Variable Star Observers (AAVSO), 55 Dale Gardens, 18 Balder Road, Douglasdale, Sandton 2191, South Africa}
\altaffiltext{7}{Astronomical Society of Southern Africa, Variable Star Section (ASSA)}
\altaffiltext{8}{Flarestar Observatory, San Gwann SGN 3160, Malta}
\altaffiltext{9}{British Astronomical Association, Variable Star Section, Standlake, Oxfordshire, UK}
\altaffiltext{10}{Madrona Peak Observatory, 4635 Shadow Grass Dr., Katy, Texas 77493, USA}
\altaffiltext{11}{American Association of Variable Star Observers (AAVSO), Rolling Hills Observatory, Clermont, Florida 34711, USA}
\altaffiltext{12}{Institute of Earth Systems, University of Malta, Msida, Malta}
\altaffiltext{13}{American Association of Variable Star Observers (AAVSO), Midland, Michigan, USA}
\altaffiltext{14}{Groupe Européen d’Observations Stellaires (GEOS), 23 Parc de Levesville, 28300 Bailleau l’Evêque, France}
\altaffiltext{15}{Bundesdeutsche Arbeitsgemeinschaft für Veränderliche Sterne (BAV), Munsterdamm 90, 12169 Berlin, Germany}
\altaffiltext{16}{Vereniging voor Sterrenkunde (VVS), Oostmeers 122 C, 8000 Brugge, Belgium}
\altaffiltext{17}{American Association of Variable Star Observers (AAVSO), New Smyrna Beach, Florida, USA}
\altaffiltext{18}{Bethune Observers Group, Post Office Box 25553 Columbia, South Carolina 29224, USA}
\altaffiltext{19}{Universidad Complutense de Madrid, Madrid, Madrid, Spain}
\altaffiltext{20}{American Association of Variable Star Observers (AAVSO), Capraia e Limite, Firenze, Italy}
\altaffiltext{21}{Variable Star Observers League in Japan (VSOLJ), 1001-105 Nishiterakata, Hachioji-shi, Tokyo 192-0153, Japan}
\altaffiltext{22}{Variable Star Observers League in Japan (VSOLJ), 7-1 Kitahatsutomi, Kamagaya-shi, Chiba 273-0126, Japan}
\altaffiltext{23}{American Association of Variable Star Observers (AAVSO), 2405 Steele Ranch Ct., Friendswood, Texas, USA}
\altaffiltext{24}{Svensk Amator Astronomisk Forening, variabelsektionen (SAAF), Sweden}
\altaffiltext{25}{URSA Astronomical Association, Variable Star Section (URSA), Finland}
\altaffiltext{26}{American Association of Variable Star Observers (AAVSO), Hillsborough, California, USA}
\altaffiltext{27}{Bronberg Observatory, Center for Backyard Astrophysics Pretoria, PO Box 11426, Tiegerpoort 0056, South Africa}
\altaffiltext{28}{Kleinkaroo Observatory, Center for Backyard Astrophysics Kleinkaroo, Sint Helena 1B, PO Box 281, Calitzdorp 6660, South Africa}
\altaffiltext{29}{Observatorio de Sencelles, Sencelles, Balearic Islands, Spain}
\altaffiltext{30}{Variable Star Observers League in Japan (VSOLJ), 290-383, Ogata-cho, Sasebo-shi, Nagasaki, Japan}
\altaffiltext{31}{Variable Star Observers League in Japan (VSOLJ), Okayama, Japan}
\altaffiltext{32}{Variable Star Observers League in Japan (VSOLJ), 5-9-3 B-305 Honson, Chigasaki-shi, Kanagawa 253-0042, Japan}
\altaffiltext{33}{Al Sadeem Astronomy, Al Wathba South, Abu Dhabi, UAE}
\altaffiltext{34}{Al Khatim Observatory, International Astronomical Center, Abu Dhabi, UAE}
\altaffiltext{35}{Agrupació Astronòmica de Sabadell, Carrer Prat de la Riba, 116, Sabadell, Barcelona 08206, Spain}
\altaffiltext{36}{Perth Observatory, Western Australia, Australia}
\altaffiltext{37}{AstroMallorca, Asociación Astronómica de Mallorca, Spain}
\altaffiltext{38}{American Association of Variable Star Observers (AAVSO), Calle San Vicente, 18, 02660 Caudete, Spain}
\altaffiltext{39}{Pobedy street, house 7, flat 60, Yuzhno-Morskoy, Nakhodka, Primorsky Krai 692954, Russia}
\altaffiltext{40}{American Association of Variable Star Observers (AAVSO), 185 Alewife Brook Parkway, Suite 410, Cambridge, Massachusetts 02138, USA}
\altaffiltext{41}{Burke-Gaffney Observatory, Saint Mary’s University, 923 Robie Street, Halifax, Nova Scotia B3H3C3, Canada}
\altaffiltext{42}{Abbey Ridge Observatory, 45 Abbey Rd, Stillwater Lake, Nova Scotia B3Z1R1, Canada}
\altaffiltext{43}{iTelescope.Net Observatory New Mexico Skies at Mayhill, New Mexico, USA}
\altaffiltext{44}{School of Physics and Astronomy, Rochester Institute of Technology, Rochester, New York 14623, USA}
\altaffiltext{45}{Observatorio Astronómico "Nastro Verde", Sorrento, Naples, Italy}
\altaffiltext{46}{Variable Star Observers League in Japan (VSOLJ), Nishi juni-jou minami 3-1-5, Nayoro-shi, Hokkaido, Japan}
\altaffiltext{47}{Observation and Data Center for Cosmosciences, Faculty of Science, Hokkaido University, Kita-ku, Sapporo-shi, Hokkaido 060-0810, Japan}
\altaffiltext{48}{Nayoro Observatory, 157-1 Nisshin, Nayoro-shi, Hokkaido 096-0066, Japan}
\altaffiltext{49}{American Association of Variable Star Observers (AAVSO), PO Box 274, Wheelock, Texas 77882, USA}
\altaffiltext{50}{Factory Physics, Inc., Houston, Texas, USA}
\altaffiltext{51}{Burleith Observatory, 1810 35th St NW, Washington, D.C. 20007, USA}
\altaffiltext{52}{Center for Backyard Astrophysics Belgium, Walhostraat 1A, B-3401 Landen, Belgium}
\altaffiltext{53}{Center for Backyard Astrophycis Extremadura, e-EyE Astronomical Complex, ES-06340 Fregenal de la Sierra, Spain}
\altaffiltext{54}{Vihorlat Observatory, Mierova 4, 06601 Humenne, Slovakia}
\altaffiltext{55}{American Association of Variable Star Observers (AAVSO), 126 Powell Bay Road, Elgin, Ontario, Canada}
\altaffiltext{56}{Royal Astronomical Society of Canada (RASC), Elgin, Ontario, Canada}
\altaffiltext{57}{Bundesdeutsche Arbeitsgemeinschaft fur Veranderliche Sterne e.V. (BAV), Oberwiesenthal, Germany}
\altaffiltext{58}{Drebach-South Observatory, Tivoli Southern Sky Guest Farm, Khomas, Namibia}
\altaffiltext{59}{American Association of Variable Star Observers (AAVSO), Glasgow, UK}

\email{mrok@kusastro.kyoto-u.ac.jp}


\maketitle

\begin{abstract}

    We present our optical photometric observations of the 2022 eruption of the recurrent nova U Scorpii (U Sco)
    using 49,152 data points over 70 d following the optical peak.
    We have also analyzed its soft X-ray (0.3--1 keV) light curve by the Neil Gehrels {\it Swift} Observatory.
    During the 2022 eruption, the optical plateau stage started 13.8--15.0 d and ended 23.8--25.0 d after the optical peak.
    The soft X-ray stage started 14.6--15.3 d and ended 38.7--39.5 d after the optical peak.
    Both stages started later and had shorter durations, 
    and the soft X-ray light curve peaked earlier and was less luminous compared to those during the U Sco 2010 eruption. 
    These points suggest that there were differences in the envelope mass between the different cycles of the nova eruption. 
    Furthermore, we have analyzed the optical eclipses during the 2022 eruption.
    The primary eclipse was first observed 10.4--11.6 d after the optical peak, 
    earlier than the beginning of the optical plateau stage.
    This sequence of events can be explained by the receding ejecta photosphere associated with the expanding nova ejecta. 
    We have determined the ingress and egress phases of the primary eclipses
    and estimated the outer radius of the optical light source centered at the white dwarf (WD).
    During the optical plateau stage, the source radius remained $\sim$1.2 times larger than the Roche volume radius of the primary WD, being close to the L1 point.
    When the optical plateau stage ended, the source radius drastically shrank to the tidal truncation radius within a few orbital periods.
    This previously unresolved phenomenon can be interpreted as a structural change in U Sco
    where the temporarily expanded accretion disk due to the nova wind returned to a steady state.
    
\end{abstract}


\section{Introduction}\label{sec01:intro}
    Novae are luminous eruptions in binaries where a white dwarf (WD) accretes hydrogen-rich material from the secondary star \citep{Nova}. 
    When the accreted layer reaches a critical mass, it undergoes a thermonuclear runaway (TNR), and a nova eruption takes place 
    (e.g., \citealp{TNR}; \citealp{chomiuk2021} for a recent review).
    Thereafter, accretion onto the WD resumes in time, and another nova eruption occurs.
    Generally speaking, the recurrence time between nova eruptions depends on the WD mass and the accretion rate 
    (\citealp{wolf2013}; see also figure 2 of \citealp{chomiuk2021}).
    Recurrent Novae (RNe) are novae whose recurrence times are short enough, 
    less than roughly 100 yr \citep{schaefer2010_1}, 
    to have more than one observed nova eruption.

    When a nova eruption takes place, the optical light curve shows a sudden brightening and reaches its peak within one day, 
    with an amplitude of $\geq$ 10 mag \citep[e.g., section 5 of][]{warner}.
    After the optical peak, it enters the early decline stage, 
    followed by the transition stage (such as the optical plateau stage as below) 
    starting at 3--4 mag and ending at $\sim$6 mag below the optical peak \citep{warner}.
    On the other hand, the soft X-ray light curve reaches its peak later than the optical one.
    
    \citet{hachisu2006_2} describe the evolution of the nova eruption in detail as follows.
    After a TNR sets in, a large part of the envelope is ejected as wind, 
    and the ejecta photosphere centered at the WD expands beyond the binary system.
    Therefore, the whole binary system is obscured in the photosphere just after the optical peak.
    Then, the photosphere lags behind the head of the ejecta and eventually begins to shrink and recede toward the WD as the ejecta density decreases.
    Finally, the optically thick WD wind gradually weakens and stops, causing the photosphere to move back to near the WD surface.
    On the surface, the remaining envelope still causes steady H-burning 
    and emits an atmospheric spectrum of several $10^5\ \mathrm{K}$.
    This steady supersoft X-ray source (SSS) leads to the detection of luminous soft X-ray photons.
    This scenario results in the soft X-ray light curve peaking later than the optical one.
    
    U Scorpii (U Sco) is one of the most well-observed RNe, and 11 nova eruptions were reported: 
    in 1863, 1906, 1917, 1936, 1945, 1969, 1979, 1987, 1999, 2010, and recently in 2022.
    The primary WD is considered to be close to the Chandrasekhar limit \citep{hachisu2000}.
    They have modeled the light curve of the 1999 eruption of U Sco and estimated the WD mass to be $1.37 \pm 0.01\ \MO$.
    The secondary star is considered to be a slightly evolved main sequence star 
    filling its Roche lobe after a large part of the central hydrogen has been consumed \citep{hachisu2000}.
    \citet{thoroughgood2001} have derived the secondary mass as $0.88 \pm 0.17\ \MO$. 
    \citet{maxwell2014} have observationally estimated that the secondary is a sub-giant of spectral type F2-G3 at the 95\% confidence level.
    In addition, its orbital inclination is calculated to be $82.7\degree \pm 2.9\degree$ 
    and high enough to show optical eclipses in quiescence \citep{thoroughgood2001}.

    Among the previous eruptions of U Sco, the 2010 eruption has been studied in great detail.
    During the U Sco 2010 eruption, its optical brightness rapidly increased from the pre-eruption magnitude $V \sim 18.0$
    and reached its peak at $V \sim 7.5$ mag within one day \citep{schaefer2010}.
    After the optical peak, it faded rapidly by three magnitudes in $2.6$ d \citep{schaefer2010}.
    When the early decline stage ended at $V \sim 14.0$ mag, it entered the optical plateau stage \citep{schaefer2010_2}, 
    which is defined as the stage when the optical magnitude temporarily remains nearly constant.
    Furthermore, U Sco gradually began to show optical eclipses almost simultaneously \citep{schaefer2011}. 
    They have precisely calculated its orbital period $P_\mathrm{orb} = 1.23054695(24)$ d and derived epochs of the middle of the primary eclipse

    \begin{equation} 
    \mathrm{BJD\ (TT):}\ 2451234.5395 (\pm 0.0005) + N \times P_\mathrm{orb}.
    \label{epoch}
    \end{equation}
    Moreover, the Neil Gehrels {\it Swift} Observatory \citep{Swift} detected U Sco in the X-ray band during the 2010 eruption \citep{schlegel2010}, 
    and its soft X-ray count rate increased rapidly $\sim$12 d after the optical peak \citep{schlegel2010_2}.
    In the case of U Sco, these soft X-ray photons were observed, not directly from the WD surface,
    but from the hot plasma around the WD via Thomson scattering (\citealp{ness2012}; \citealp{orio2013}; \citealp{mkato2020}). 
    It should be noted that epochs during the 2010 eruption are all described in Heliocentric Julian Date (HJD) in \citet{schaefer2011}.
    However, in this paper, we describe all observation times including the 2010 eruption   
    in the Barycentric Julian Date based on the Terrestrial Time (BJD (TT)). 
    BJD (TT) is a uniform time system in an inertial reference frame and not affected by leap seconds within the time span from 2010 to 2022,
    allowing us to consistently treat the epochs for the long time span.

    \citet{schaefer2011} describe these phenomena in detail, particularly focusing on the optical plateau stage and the optical eclipses with numerous data points.
    However, only a few novae have been studied for differences in their optical and soft X-ray light curves between different eruptions occurring in the same object.
    Moreover, \citet{hachisu2000} have suggested that the optical plateau stage is mainly attributed to the accretion disk irradiated by the WD,
    based on the theoretical light-curve model for the U Sco 1999 eruption.
    They have also suggested the possibility of the accretion disk expanding to become larger during the optical plateau stage than in quiescence.
    However, there have been few observational studies tracking the process of structural changes in the accretion disk that would support this idea.

    The 2022 eruption of U Sco was firstly reported by Masayuki Moriyama on 2022 June 6.72 UTC 
    (BJD (TT) 2459737.23; vsnet-alert 26798\footnote{http://ooruri.kusastro.kyoto-u.ac.jp/mailarchive/vsnet-alert/26798}).
    We successfully obtained rich and detailed optical photometric data points during the U Sco 2022 eruption compared to those during the previous ones.
    As for soft X-ray observations, {\it Swift} detected the beginning of the SSS stage, as reported by \citet{page2022}.
    Subsequently, \citet{page2022_2} have reported an increase in X-ray count rate and its peak after the eruption.
    The X-ray light curve and its maximum count rate are described in \citet{evans2023}.

    Our purpose in this paper is to present our optical photometry of the U Sco 2022 eruption, 
    and to conduct a comparative study between the 2010 and 2022 eruptions.
    We aim to examine whether the following three phenomena vary between the different eruptions of U Sco:
    the optical plateau stage, soft X-ray stage, and optical eclipses. 
    To assess the different behavior of the soft X-ray light curve accurately,
    we adopt a unified photometry method using {\it Swift} archival data for both eruptions.
    Additionally, by estimating the outer radius of the optical light source centered at the primary WD, based on the eclipse width, 
    we aim to confirm that the optical light source during the optical plateau stage is mainly the accretion disk, 
    and to examine its structural changes.
    In section \ref{sec02:obs}, we describe optical and soft X-ray photometry. 
    Our results on the optical plateau stage, soft X-ray stage, and optical eclipses are presented in section \ref{sec03:res}.
    Discussion and summary follow in sections \ref{sec04:dis} and \ref{sec05:sum}, respectively.

\section{Observations} \label{sec02:obs}
    In this section, we describe our optical photometry of the U Sco 2022 eruption, 
    as well as soft X-ray photometry of both the 2022 and 2010 eruptions.
    In this paper, we also use optical photometry of the 2010 eruption as a comparison, which is described in detail in \citet{schaefer2011}.

\subsection{Optical photometry} \label{subsec02:opt}
    \begin{table}[tb]
        \tbl{List of observers and instruments for optical photometry.\footnotemark[$*$]}{%
        \begin{tabular}{cc}
           \hline
            Code & Observer (Observatory) / Telescope \& CCD  \\ 
            \hline
            HaC & Franz-Josef Hambsch / 40 cm telescope and FLI ML16803 \\
            KU1 & Kyoto Univ. Campus Obs. / 40 cm SCT and Apogee U6 \\
            MLF & Berto Monard (CBA Kleinkaroo) / 30 cm RCX400 and SBIG ST8 \\
            \hline
        \end{tabular}}\label{tab02:obs}
        \begin{tabnote}
            \footnotemark[$*$] This is just a sample of e-table 1. \\
        \end{tabnote}
    \end{table}
    
    \begin{table}[tb]
        \tbl{Log of optical photometry of the U Sco 2022 eruption.\footnotemark[$*$]}{%
        \begin{tabular}{rrrrrr}
           \hline
            Start\footnotemark[$\dagger$] & End\footnotemark[$\dagger$] & Mag\footnotemark[$\ddagger$] & $\sigma_{\mathrm{mag}}$\footnotemark[$\S$] &
            $N$\footnotemark[$\|$] & Code\footnotemark[$\#$] \\ 
            \hline
            59738.043 & 59738.202 & 8.209 & 0.097 & 1247 & KU1 \\ 
            59743.557 & 59743.847 & 11.470 & 0.076 & 1271 & HaC \\ 
            59744.554 & 59744.844 & 12.201 & 0.065 & 969 & HaC \\
            59745.476 & 59745.841 & 12.573 & 0.107 & 1693 & HaC \\ 
            59746.225 & 59746.576 & 12.587 & 0.078 & 994 & MLF \\
            \hline
        \end{tabular}}\label{tab02:opt}
        \begin{tabnote}
            \footnotemark[$*$] This is just a sample of e-table 2. \\
            \footnotemark[$\dagger$] BJD (TT) $-$ 2400000. \\
            \footnotemark[$\ddagger$] Mean magnitude. \\
            \footnotemark[$\S$] Standard deviation of the magnitudes. \\
            \footnotemark[$\|$] Number of observations. \\
            \footnotemark[$\#$] Observer's code (see table \ref{tab02:obs}).
        \end{tabnote}
    \end{table}
    
    After the first report on the U Sco 2022 eruption as mentioned above, 
    time-resolved CCD photometric observations in optical had been performed 
    by the Variable Star Network collaboration \citep[VSNET;][]{VSNET}\footnote{http://www.kusastro.kyoto-u.ac.jp/vsnet/}
    and the Variable Star Observers League in Japan (VSOLJ).\footnote{http://vsolj.cetus-net.org/index.html}
    Data reduction and calibration with the comparison stars were performed by each observer.
    In addition to our data from VSNET and VSOLJ, we also retrieved archival data from
    the American Association of Variable Star Observers (AAVSO)\footnote{https://www.aavso.org} and
    the ASAS-SN Sky Patrol \citep{ASN_1,ASN_2}.
    One can also see an example of the optical light curve from the AAVSO archive in figure 1 of \citet{evans2023}.
    Observatories (observers) and instruments are summarized 
    in table \ref{tab02:obs}.\footnote{A complete listing of table \ref{tab02:obs} is available as e-table 1 in the supplementary data section.}
    The log of this optical photometry is summarized 
    in table \ref{tab02:opt}.\footnote{A complete listing of table \ref{tab02:opt} is available as e-table 2 in the supplementary data section.}
    As a result, we obtained 49,152 optical photometric data points during the U Sco 2022 eruption.
    It should be noted that the optical photometric data had been taken by various filters and observers, 
    so we applied zero-point corrections as follows.
    Each magnitude scale observed by each filter and observer was adjusted to the data observed by Franz-Josef Hambsch's $V$ filter  
    as he had contributed a substantial number of data points.
    
    \begin{table}[tb]
        \tbl{List of magnitudes in quiescence.}{%
        \begin{tabular}{ccccc}
           \hline
            Band & Mag\footnotemark[$*$] & $\sigma$\footnotemark[$\dagger$] & $N$\footnotemark[$\ddagger$] & Observatory \\ 
            \hline
            $g$ & 18.3 & 0.228 & 25 & ZTF \\
            $r$ & 18.2 & 0.200 & 32 & ZTF \\
            $o$ (orange) & 18.6 & 0.662 & 49 & ATLAS \\
            $c$ (cyan) & 18.8 & 0.417 & 14 & ATLAS \\
            \hline
        \end{tabular}}\label{tab02:quie}
        \begin{tabnote}
            \footnotemark[$*$] Mean magnitude. \\
            \footnotemark[$\dagger$] Standard deviation of the magnitudes. \\
            \footnotemark[$\ddagger$] Number of observations. \\
        \end{tabnote}
    \end{table}
    
    To cover quiescence, we also retrieved archival data during the last two years before the U Sco 2022 eruption 
    from the Zwicky Transient Facility \citep[ZTF;][]{ZTF} and
    the Asteroid Terrestrial-impact Last Alert System \citep[ATLAS;][]{ATLAS}.
    The magnitude scale was also adjusted to the data observed by Franz-Josef Hambsch's $V$ filter.
    It should be noted that we excluded data during the eclipses for calculating the mean and deviation of the observed magnitudes.
    As a result, we obtained 120 pre-eruption data points, which are summarized in table \ref{tab02:quie}.

\subsection{Soft X-ray photometry} \label{subsec02:softX}
    \begin{table}[tb]
        \tbl{Log of soft X-ray photometry of the U Sco 2022 eruption.\footnotemark[$*$]}{%
        \begin{tabular}{llr}
           \hline
             ObsID & Start (UTC) & Exposure time (s) \\ 
            \hline
            00031417043 & 2022-06-07 04:33:34 & 972.82900 \\
            00031417044 & 2022-06-07 07:49:36 & 850.08100 \\
            00031417045 & 2022-06-07 10:54:34 & 995.39700 \\
            00031417046 & 2022-06-07 14:17:35 & 965.30800 \\
            00031417047 & 2022-06-07 23:38:36 & 273.28800 \\       
            \hline
        \end{tabular}}\label{tab02:softx2022}
        \begin{tabnote}
            \footnotemark[$*$] This is just a sample of e-table 3. \\
        \end{tabnote}
    \end{table}

    \begin{table}[tb]
        \tbl{Log of soft X-ray photometry of the U Sco 2010 eruption.\footnotemark[$*$]}{%
        \begin{tabular}{llr}
           \hline
             ObsID & Start (UTC) & Exposure time (s) \\ 
            \hline
            00031417003 & 2010-01-29 01:03:00 & 1762.79000 \\
            00031417004 & 2010-01-29 01:06:00 & 14296.41100 \\
            00031417005 & 2010-01-31 01:07:00 & 1573.06400 \\
            00031417006 & 2010-01-31 01:10:00 & 16897.09400 \\
            00031417007 & 2010-02-02 01:29:00 & 1956.20300 \\
            \hline
        \end{tabular}}\label{tab02:softx2010}
        \begin{tabnote}
            \footnotemark[$*$] This is just a sample of e-table 4. \\
        \end{tabnote}
    \end{table}
    
    We obtained numerous archival data 
    from the Neil Gehrels {\it Swift} satellite X-ray telescope \citep[{\it Swift}-XRT;][]{XRT} in Photon Counting mode.
    Event files were calibrated and cleaned 
    with the {\it Swift}-XRT software task {\it xrtpipeline} distributed with the HEASoft package.\footnote{http://heasarc.gsfc.nasa.gov/lheasoft/}
    In this paper, we defined the energy band for soft X-ray as $0.3$--$1$ keV.
    The source region were designed to be a circle of radius 0.01 degrees centered at $(\alpha, \delta)_{\rm J2000.0} =$ (\timeform{16h22m30s.5}, $-$\timeform{17D52'43".2}),
    and we obtained 105 and 196 soft X-ray photometric data points for the U Sco 2022 and 2010 eruptions, respectively. 
    The soft X-ray data are in units of counts per second and binned to 0.1 d.
    The logs of soft X-ray photometry of the 2022 and 2010 eruptions are summarized 
    in table \ref{tab02:softx2022},\footnote{A complete listing of table \ref{tab02:softx2022} is available as e-table 3 in the supplementary data section.}
    and in table \ref{tab02:softx2010},\footnote{A complete listing of table \ref{tab02:softx2010} is available as e-table 4 in the supplementary data section.}
    respectively.

\section{Results} \label{sec03:res}

\subsection{Optical light curve} \label{subsec03:pla}

    \begin{figure*}[tb]
        \begin{center}
            \includegraphics[width=\linewidth]{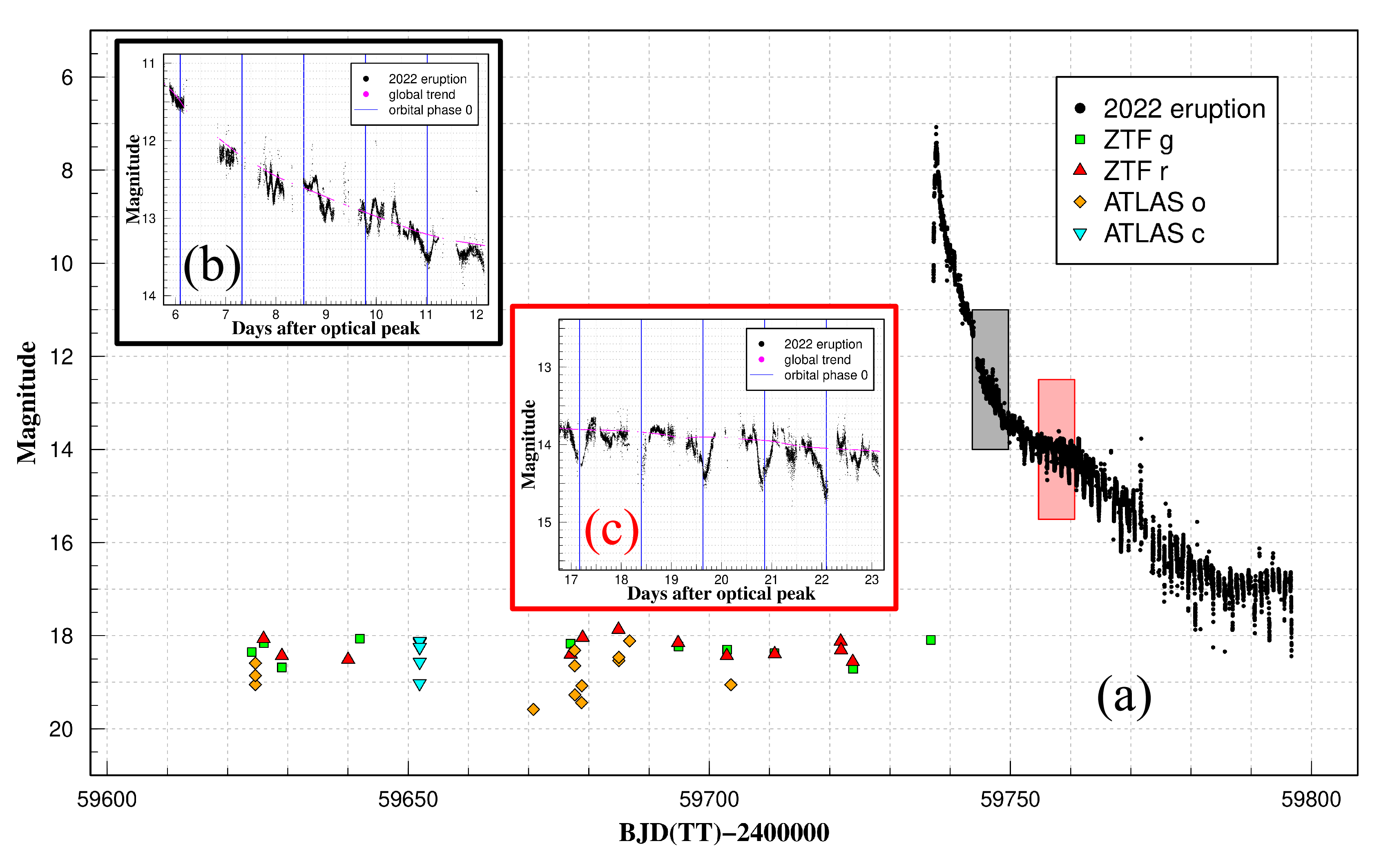} 
        \end{center}
        \caption{(a) Overall optical light curve of the U Sco 2022 eruption (black filled circles). 
        Data are binned to 0.001 d. 
        Magnitudes in quiescence are also plotted partially 
        (green filled squares for the ZTF $g$ band, red filled triangles for the ZTF $r$ band,
        orange filled diamonds for the ATLAS $o$ band, and cyan filled inverted triangles for the ATLAS $c$ band).
        (b) Enlarged optical light curve of the 2022 eruption between day 6 and day 12 after the optical peak (black filled circles).
        It is the enlarged view of the black shaded box in panel (a).
        Magenta filled circles represent the global trend of the 2022 eruption calculated by LOWESS (see Appendix 1).
        Blue lines represent epochs of the middle of the primary eclipse, namely, the orbital phase 0, calculated by equation (\ref{epoch}).
        (c) Enlarged optical light curve of the 2022 eruption between day 17 and day 23 after the optical peak.
        It is the enlarged view of the red shaded box in panel (a).}
        \label{fig03:lc_2022}
    \end{figure*}
    
    Figure \ref{fig03:lc_2022}(a) shows the optical light curve of the U Sco 2022 eruption (black filled circles). 
    The optical peak magnitude during the 2022 eruption was approximately 7.5 mag on BJD (TT) 2459737.68(3), 
    based on the global trend calculated by LOWESS (see Appendix 1).
    The last detection before the 2022 eruption was approximately 18.1 mag in the ZTF $g$ band on BJD (TT) 2459736.80, 
    which indicates that the optical light curve reached its peak within one day following the nova eruption.  

    \begin{figure}[tb]
        \begin{center}
            \includegraphics[width=\linewidth]{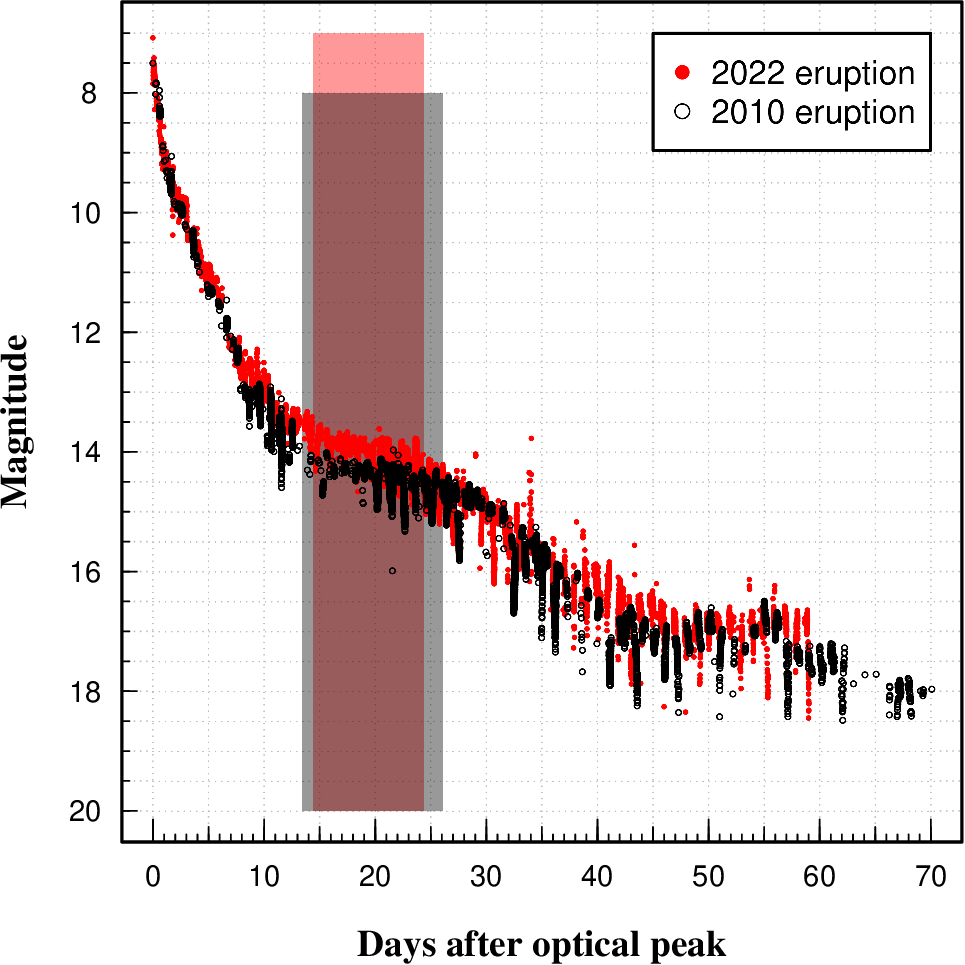}
        \end{center}
        \caption{Optical light curves of the U Sco 2022 (red filled circles) and 2010 (black open circles) eruptions. 
        The red and black regions represent the optical plateau stages for the 2022 (red) and 2010 (black) eruptions, respectively.
        The optical peak magnitude during the 2010 eruption was on BJD (TT) 2455224.69(7) \citep{schaefer2010}.
        The 2010 eruption data are derived from \citet{schaefer2011} and VSOLJ.}\label{fig03:plateau}
    \end{figure}

    Figure \ref{fig03:plateau} shows the optical light curves of both the U Sco 2022 (red) and 2010 (black) eruptions.
    While fading, both eruptions exhibited the optical plateau stage, 
    when the optical magnitude remained nearly constant except for variations caused by the eclipses.
    We calculated its beginning and ending dates as follows.
    We first roughly estimated the optical plateau interval.
    Then, we created a histogram of the out-of-eclipse data points with a binsize of 0.1 mag during this assumed interval for each eruption.
    Subsequently, we measured a range of the out-of-eclipse magnitude during the optical plateau stage, 
    based on the partitions showing high frequency in the histogram.
    The results were 13.8--14.0 mag and 14.1--14.3 mag for the 2022 and 2010 eruptions, respectively.
    Finally, we estimated when the out-of-eclipse magnitude fell within these ranges with an uncertainty of around one orbital period. 

    In the 2022 eruption, the optical plateau stage started 13.8--15.0 d and ended 23.8--25.0 d after the optical peak, 
    which resulted in a duration of 8.8--11.2 d.
    In the 2010 eruption, the optical plateau stage started 12.8--14.0 d and ended 25.5--26.7 d after the optical peak, 
    which resulted in a duration of 11.5--13.9 d. 
    As a result, it turns out that the optical plateau stage started later and had a shorter duration in the 2022 eruption than that in the 2010 eruption, 
    while the optical plateau magnitude was $\sim$0.3 mag brighter in the 2022 eruption.

\subsection{Soft X-ray light curve} \label{subsec03:softX}
    \begin{figure}[tb]
        \begin{center}
            \includegraphics[width=\linewidth]{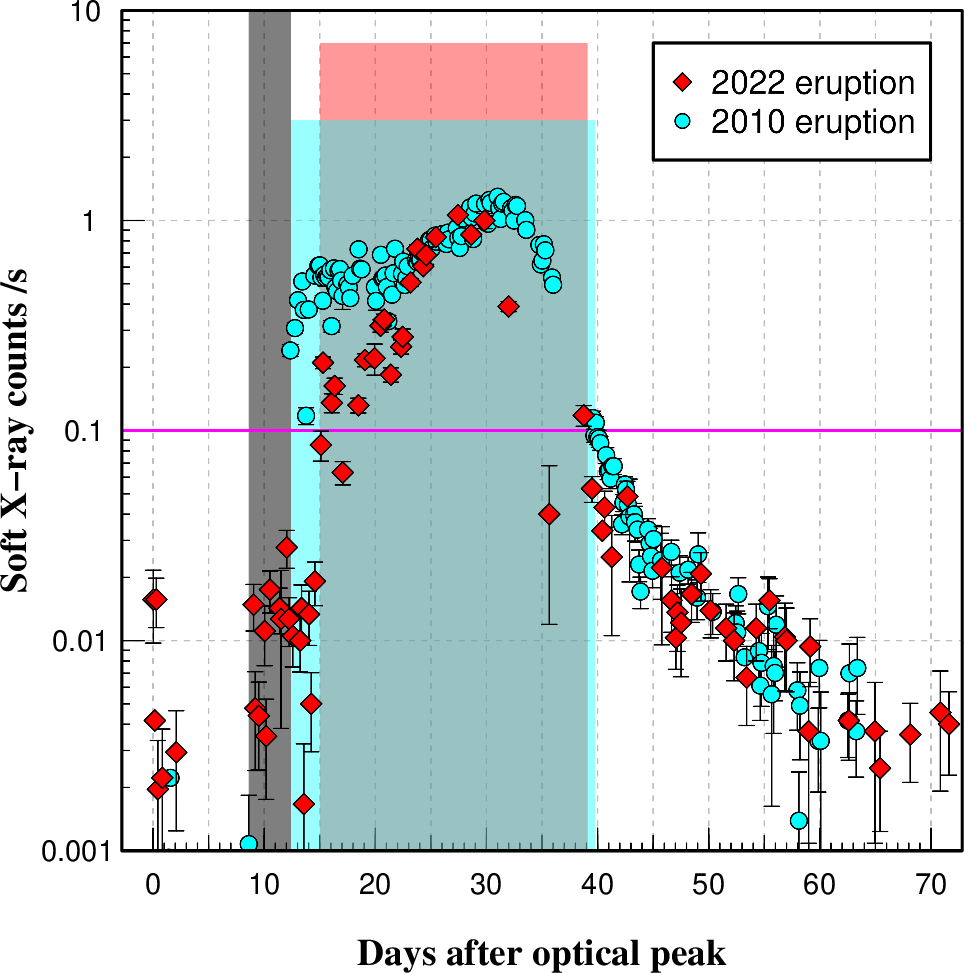}
        \end{center}
        \caption{Soft X-ray (0.3-1 keV) light curves of the U Sco 2022 (red filled diamonds) and 2010 (cyan filled circles) eruptions.
        The red and cyan regions represent the soft X-ray stages for the 2022 (red) and 2010 (cyan) eruptions, respectively.
        The gray region represents the interval of no observations during the 2010 eruption.
        The magenta line represents a threshold for the beginning and ending of the soft X-ray stage.}\label{fig03:softx}
    \end{figure}

    Figure \ref{fig03:softx} shows the soft X-ray (0.3--1 keV) light curves of both the U Sco 2022 (red) and 2010 (cyan) eruptions.
    It is clear that both soft X-ray light curves reached their peaks at the value $\sim$1 count per second, 
    and $\sim$30 d later than the optical ones in both eruptions. 
    Again, we defined the soft X-ray stage, when the soft X-ray photons were evident, and calculated its beginning and ending dates for each eruption.
    \citet{mkato2020} defined the X-ray turn-on time (turn-off time) as the time when the X-ray flux rises (decreases) to a tenth of its peak value.  
    Therefore, we established a threshold for the beginning and ending of the soft X-ray stage at the value 0.1 counts per second (magenta line in figure \ref{fig03:softx}).
    
    In the 2022 eruption, the soft X-ray stage started 14.6--15.3 d and ended 38.7--39.5 d after the optical peak,
    which resulted in a duration of 23.4--24.9 d.
    The soft X-ray light curve peaked 27.4--29.8 d after the optical peak at the value 0.96--1.06 counts per second.
    In the 2010 eruption, the soft X-ray stage started 8.6--12.4 d and ended 39.6--40.0 d after the optical peak,
    which resulted in a duration of 27.2--31.4 d.
    The soft X-ray light curve peaked 29.1--32.7 d after the optical peak at the value 1.20--1.30 counts per second.
    It should be noted that there were no soft X-ray observations by {\it Swift} during days 8.6--12.4 in the 2010 eruption (gray region in figure \ref{fig03:softx}), 
    so we conservatively determined the beginning date of the soft X-ray stage in 2010.
    As a result, it turns out that the soft X-ray stage started later, peaked earlier, had a shorter duration, and was less luminous 
    in the 2022 eruption than that in the 2010 eruption.

    \subsection{Optical eclipse}
    \begin{figure}[tb]
        \begin{center}
            \includegraphics[width=0.85\linewidth]{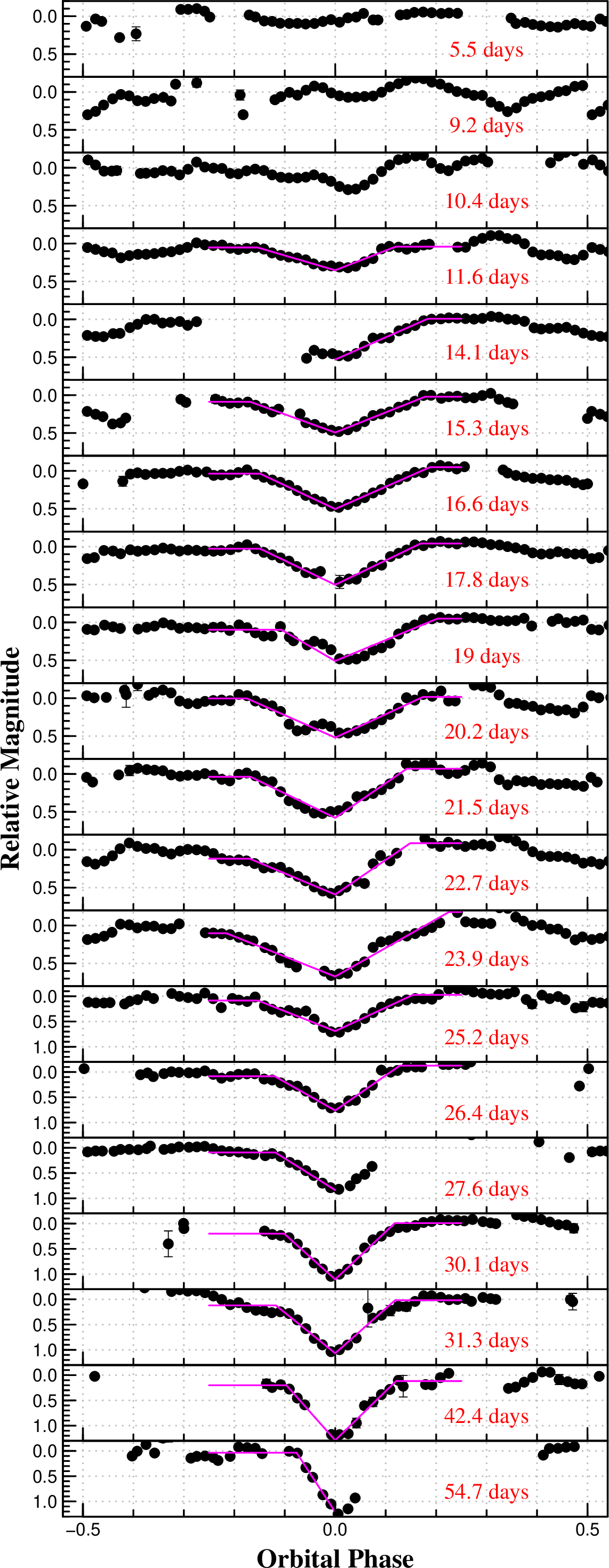}
        \end{center}
        \caption{Phase-averaged profiles during the U Sco 2022 eruption.
        Red labels represent the central times of the sampling intervals in days after the optical peak.
        Magenta lines represent the regression models with the determined parameters (see also figure \ref{figapp:mcmc}).
        Note that each sampling interval is two epochs, centered at the orbital phase 0.5 (see Appendix 1).}\label{fig03:phv}
    \end{figure}

    Figure \ref{fig03:lc_2022}(b) shows the optical light curve 
    between day 6 and day 12 after the optical peak during the U Sco 2022 eruption in detail (black filled circles).
    Between day 7 and day 10, short-term variations were clearly present at any orbital phases.
    On day around 11, however, these optical flickerings were much less presented, and instead, 
    we distinctly detected a local minimum at the orbital phase 0 on BJD (TT) 2459748.69 (blue line in figure \ref{fig03:lc_2022}(b)).
    
    Figure \ref{fig03:phv} shows some excerpts of the phase-averaged profiles during the U Sco 2022 eruption (see Appendix 1).
    Relative magnitudes, which were obtained by subtracting the global trend from the observed data points, are folded at the orbital period. 
    Epochs of the expected primary eclipse correspond to the orbital phase 0.
    
    Based on these profiles, any variations due to the eclipses were not detected immediately after the optical peak.
    However, a shallow dip was observed on day 10.4 after the optical peak (3rd panel from the top in figure \ref{fig03:phv}),  
    and on day 11.6 (4th panel in figure \ref{fig03:phv}), a minimum of the relative magnitude was clearly detected at the orbital phase 0, 
    indicating the emergence of the primary eclipse. 
    Therefore, it is estimated that the primary eclipse became detectable 10.4-11.6 d after the optical peak during the 2022 eruption.
    The time sequence of the optical plateau stage, soft X-ray stage, and optical eclipses in this section are listed in table \ref{tab03:value}.

    \begin{table}[tb]
        \tbl{Time sequence of the optical plateau stage, soft X-ray stage, and optical eclipses in the U Sco 2022 and 2010 eruptions.}{%
        \begin{tabular}{lcc}
            \hline
            Event & 2022 eruption & 2010 eruption \\ 
            \hline
            Optical plateau stage & & \\ 
            \ \ \ \ \ beginning date\footnotemark[$*$] & 13.8--15.0 & 12.8--14.0 \\ 
            \ \ \ \ \ ending date\footnotemark[$*$] & 23.8--25.0 & 25.5--26.7 \\ 
            \ \ \ \ \ duration (d) & 8.8--11.2 & 11.5--13.9 \\
            \hline
            Soft X-ray stage & & \\ 
            \ \ \ \ \ beginning date\footnotemark[$*$] & 14.6--15.3 & 8.6--12.4 \\ 
            \ \ \ \ \ ending date\footnotemark[$*$] & 38.7--39.5 & 39.6--40.0 \\ 
            \ \ \ \ \ duration (d) & 23.4--24.9 & 27.2--31.4 \\
            \hline
            The primary eclipse &  & \\
            \ \ \ \ \ first observed date\footnotemark[$*$] & 10.4--11.6 & $+$10.2\footnotemark[$\dagger$] \\
            \hline
        \end{tabular}}\label{tab03:value}
        \begin{tabnote}
            \footnotemark[$*$] Days after the optical peak. \\
            \footnotemark[$\dagger$] Reanalysis of the data in \citet{schaefer2011} by \citet{mason2012}.
        \end{tabnote}
    \end{table}

\subsection{Eclipse width}\label{subsec04:width}
    \begin{table}[tb]
        \tbl{List of values for ingress and egress phases of the primary eclipse during the U Sco 2022 eruption.}{%
        \begin{tabular}{ccccc}
            \hline
            Date\footnotemark[$*$] & Ingress phase & $\alpha_{\mathrm{ing}}$\footnotemark[$\dagger$] & 
            Egress phase & $\alpha_{\mathrm{eg}}$\footnotemark[$\ddagger$] \\ 
            \hline
            11.6 & $-$0.1539(34) & 1.077(23) & 0.1199(28) & 0.796(28) \\
            14.1 & -             & -         & 0.1837(14) & 1.228(5)  \\ 
            15.3 & $-$0.1683(27) & 1.162(14) & 0.1780(20) & 1.206(8)  \\
            16.6 & $-$0.1488(20) & 1.042(14) & 0.1878(22) & 1.241(6)  \\
            17.8 & $-$0.1488(18) & 1.042(13) & 0.1693(18) & 1.167(9)  \\
            19.0 & $-$0.1065(15) & 0.657(16) & 0.2027(15) & 1.270(2)  \\
            20.2 & $-$0.1746(11) & 1.192(5)  & 0.1740(11) & 1.189(5)  \\
            21.5 & $-$0.1678(14) & 1.159(7)  & 0.1419(10) & 0.990(8)  \\
            22.7 & $-$0.1697(18) & 1.169(9)  & 0.1487(14) & 1.041(10) \\
            23.9 & $-$0.2158(22) & 1.276(3)  & 0.2255(14) & 1.291(2)  \\
            25.2 & $-$0.1531(17) & 1.071(13) & 0.1540(15) & 1.077(10) \\
            26.4 & $-$0.1200(12) & 0.797(12) & 0.1280(9)  & 0.873(8)  \\
            27.6 & $-$0.1187(11) & 0.785(11) & -          & -         \\
            30.1 & $-$0.1011(9)  & 0.597(10) & 0.1182(7)  & 0.780(7)  \\
            31.3 & $-$0.1176(7)  & 0.774(8)  & 0.1186(6)  & 0.784(6)  \\
            42.4 & $-$0.0967(6)  & 0.546(7)  & 0.1207(7)  & 0.804(7)  \\ 
            54.7 & $-$0.0766(5)  & 0.301(3)  & -          & -         \\ 
            \hline
        \end{tabular}}\label{tab03:mcmc}
        \begin{tabnote}
            \footnotemark[$*$] Central time of each sampling interval for each phase-averaged profile in days after the optical peak. \\
            \footnotemark[$\dagger$] Parameter $\alpha$, light-source radius normalized by the primary's Roche volume radius, calculated by ingress phase at the 99\% confidence interval. \\
            \footnotemark[$\ddagger$] Parameter $\alpha$ calculated by egress phase at the 99\% confidence interval.\\
        \end{tabnote}
    \end{table}  

    \begin{figure*}[tb]
        \begin{center}
            \includegraphics[width=\linewidth]{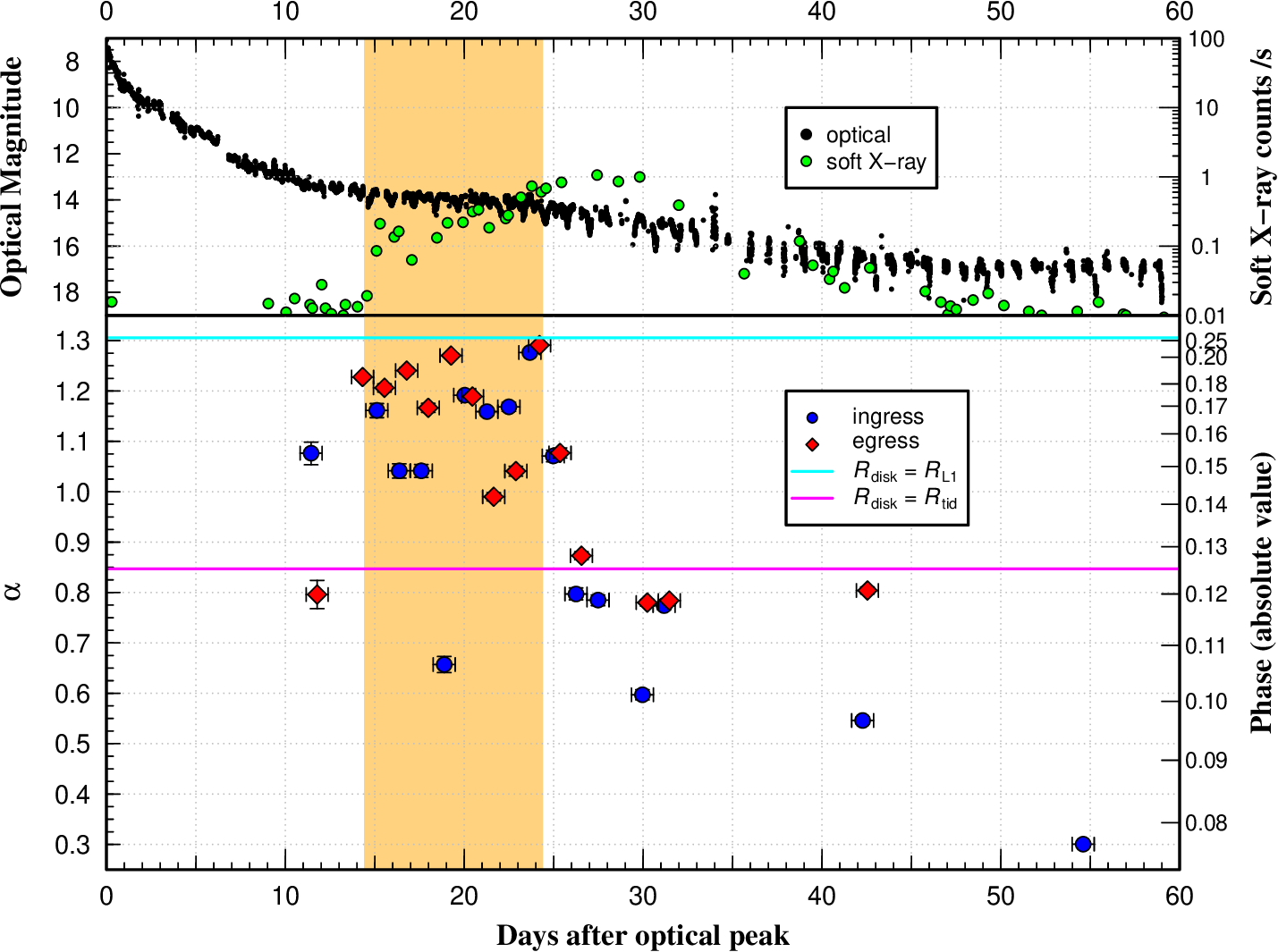} 
        \end{center}
        \caption{Upper panel: Optical (black filled circles) and soft X-ray (green filled circles) light curves of the U Sco 2022 eruption.
        Lower panel: Transition of $\alpha$, the optical light-source radius normalized by the primary's Roche volume radius (see Appendix 3),
        based on the ingress (blue filled circles) and egress (red filled diamonds) phases of the eclipses determined by the MCMC method.
        The cyan line represents the light-source radius when expanding to the distance between the L1 point and the primary WD $R_{\mathrm{L1}}$.
        The magenta line represents the light-source radius when expanding to the tidal truncation radius $R_{\mathrm{tid}}$.
        The orange region represents the optical plateau stage.}\label{fig03:ecl}
    \end{figure*}  

    One can notice a decreasing trend of the eclipse width over the eruption in figure \ref{fig03:phv}.
    To quantitatively measure the eclipse width and its evolution,
    we determined the ingress and egress phases of the observed eclipses as follows.
    Each eclipse profile was fit using a V-shaped straight line in magnitude scale,
    bending at the ingress phase, the orbital phase 0, and the egress phase along the orbital-phase axis,
    and these fitting parameters were calculated using the MCMC method (see Appendix 2).
    Magenta lines in figure \ref{fig03:phv} show the results of parameter estimation for several phase-averaged profiles during the U Sco 2022 eruption.
    It should be noted that MCMC analysis was applied to the phase-averaged profile covering a sufficient range of the orbital phase around 0.
    On day 14.1 after the optical peak (5th panel in figure \ref{fig03:phv}), 
    we only fit the observed data for orbital phases greater than 0 and determined only the egress phase.
    On days 27.6 and 54.7 (16th and 20th panels in figure \ref{fig03:phv}, respectively),
    we only fit the observed data for orbital phases less than 0 and determined only the ingress phase.

    For a further interpretation of these ingress and egress phases,
    we calculated the corresponding light-source radius $\alpha$, 
    centered at the primary WD and normalized by the primary's Roche volume radius $R_1^{*}$ (see Appendix 3).
    Table \ref{tab03:mcmc} shows these values.
    Figure \ref{fig03:ecl} shows the transition of $\alpha$ during the 2022 eruption (lower panel), 
    with optical (black) and soft X-ray (green) light curves (upper panel).
    Here, the terminology "Roche volume radius" refers to an average radius 
    yielding the same volume as the non-spherical Roche lobes (e.g., section 4.4 of \citealp{frank2002}; \citealp{leahy2015}).
    By definition, a sphere of radius $R_1^{*}$ has the same volume as the region surrounded by the primary's Roche lobe, 
    which is a teardrop-shaped region and has its apex at the L1 point 
    (see also the orange line and the black dashed line in figure \ref{figapp:roche}).
    Therefore, the axisymmetric light source centered at the primary WD first touches the L1 point
    when its outer radius expands to become $\sim$1.3 times larger than the primary's Roche volume radius 
    (cyan line in figure \ref{fig03:ecl}; see also the cyan line in figure \ref{figapp:roche}).
    During the optical plateau stage, the source radius remained $\alpha \sim 1.2$.
    When the optical plateau stage ended, the source radius drastically shrank to $\alpha \sim 0.85$ within a few orbital periods.

\section{Discussion}\label{sec04:dis}
\subsection{Optical plateau and soft X-ray stages}\label{subsec04:01}
    Both the optical plateau and soft X-ray stages started later and had shorter durations in the U Sco 2022 eruption than those in the U Sco 2010 eruption.
    Moreover, the soft X-ray light curve in the 2022 eruption peaked earlier and was less luminous compared to that in the 2010 eruption.
    One interpretation is the different envelope mass between the different nova eruptions, 
    especially the hydrogen content in the envelope,
    where convection and element mixing occur just after a TNR sets in.
    This envelope corresponds to the whole accreted hydrogen-rich layer and a portion of the processed helium-rich layer below
    \citep[see figure 1(ii) of][]{fujimoto1992}.
    
    \citet{hachisu2006} have conducted model calculations of the soft X-ray light curve of the nova eruption and 
    stated that the evolution of the nova eruption depends both on the WD mass and on the chemical composition of the envelope.
    They have calculated turn-on and turn-off times of the soft X-ray stage theoretically, 
    correlating with the hydrogen mass fraction $X$ of the envelope.
    Their results show that the turn-on time becomes later for larger hydrogen content $X$.
    Taking into account the effect of heat exchange between the H-burning zone and helium-rich layer below during the nova eruption, 
    \citet{hachisu2007} have demonstrated that the optical plateau and soft X-ray stages start later and have shorter durations when $X$ increases.
    They have also indicated that the peak soft X-ray flux decreases as $X$ becomes larger.

    For U Sco, the last three observed eruptions are in 1999 February, 2010 January, and 2022 June at optical peaks,
    indicating that the quiescent interval before the 2022 eruption is longer than that before the 2010 eruption by around one year.
    Therefore, a larger amount of hydrogen-rich material was supposed to be accreted onto the WD, 
    resulting in more massive envelope, and hence more massive ejecta during the 2022 eruption.
    This corresponds to a larger value of $X$ when a TNR sets in and 
    convection extensively mixes the hydrogen-rich accretion gas with the helium-rich layer below.
    It should be noted that there might be a missed eruption of $2016.78 \pm 0.10$ between the 2010 and 2022 eruptions as suggested by \citet{schaefer2022}.
    If this eruption had really happened, the argument in this subsection would not hold.
    
    The beginning of the soft X-ray stage can be interpreted as the time 
    when the ejecta photosphere recedes and the WD surface becomes optically thin after the nova eruption.
    Due to a larger amount of ejecta during the 2022 eruption compared to that during the 2010 eruption,
    the recession of the photosphere might take a longer time, 
    which resulted in the later beginning date of the soft X-ray stage.
    The ending of the soft X-ray stage can be interpreted as the time
    when the remaining envelope stops steady H-burning on the WD surface.
    Judging from the result of \citet{hachisu2007}, 
    due to the larger hydrogen content of the envelope during the 2022 eruption compared to that during the 2010 eruption, 
    heat flux from the helium-rich layer to the remaining H-burning envelope was reduced, 
    making it difficult to maintain a temperature high enough to emit soft X-ray photons sufficiently.
    This scenario can result in the smaller peak value, earlier peak time, and shorter duration of the soft X-ray stage during the 2022 eruption. 

    There are other comparative studies of nova eruptions occurring in the same object.
    M31N 2008-12a showed the same behavior as U Sco, while RS Oph showed the opposite behavior.
    \citet{henze2018} have conducted one on the X-ray light curves of the 2013, 2014, 2015, and 2016 nova eruptions in M31N 2008-12a, 
    which has a mean recurrence time of around one year.
    The quiescent interval before the 2016 eruption is around 35\% longer compared to those before the other three eruptions.  
    They have demonstrated that the 2016 eruption had a shorter and less luminous SSS emission, which is consistent with our observations of U Sco. 
    Meanwhile, \citet{page2022_3} have examined the differences in the optical and soft X-ray light curves between the 2006 and 2021 nova eruptions in RS Oph.
    They have demonstrated that the optical light curves of both eruptions in RS Oph were almost identical, 
    although the quiescent intervals were different.
    The 2021 eruption occurred 15.5 yr after the 2006 eruption, and the 2006 eruption occurred 21 yr after the 1985 eruption.
    They have pointed out that
    the SSS emission rose later, peaking at a lower level, and then started fading earlier in the 2021 eruption than that in the 2006 eruption,
    which is the opposite behaviour from U Sco and M31N 2008-12a.
    \citet{page2022_3} have also suggested that
    there was a slight brightening in the optical during the last 5 yr before the 2006 eruption,
    possibly indicative of a change in the accretion disk. 

    What we need to care about is that the orbital period of RS Oph is 453.6(4) d \citep{brandi2009},
    longer than the timescale of the evolution of its nova eruption $\sim$80 d. 
    Thus, we observed the SSS emission of the RS Oph 2021 eruption at the orbital phase nearly 180 degrees 
    different from that of the RS Oph 2006 eruption \citep{ness2023}.
    Soft X-ray photons, originating from the WD surface (SSS) 
    and being scattered in the electron-rich environment around the WD via Thomson scattering,
    are also absorbed by the cool plasma of the intervening gas \citep{ness2012}.
    The optical depth of low-ionization plasma or neutral hydrogen can be much higher in soft X-ray than in optical light.
    Consequently, \citet{ness2023} have concluded that the intrinsic SSS emission was the same during both RS Oph eruptions, 
    and that the different behavior of the observed SSS emission was due to viewing it at different angles, namely orbital phases, 
    through an inhomogeneous density distribution of the lower-temperature ejecta and the pre-existing stellar wind.
    They have suggested that the observed lower SSS emission in the 2021 eruption was attributed to 
    the higher line-of-sight absorption by the lower-temperature plasma being more opaque to soft X-ray.

    On the contrary, the orbital period of U Sco is $\sim$1.2 d,
    much shorter than the timescale of the evolution of its nova eruption $\sim$70 d.
    Thus, the observations of U Sco in soft X-ray covered various orbital phases, 
    unlike those of RS Oph at the more or less fixed orbital phase. 
    Although the inhomogeneous ejecta could be present in U Sco as well, 
    the trend of the soft X-ray light curve should be more affected by the overall structure of the ejecta 
    rather than the phase-dependent density profile.
    Moreover, the line-of-sight absorption was nearly comparable during both U Sco eruptions,
    with a hydrogen column density of $N(\mathrm{H}) \sim 3 \times\ 10^{21}\ \mathrm{cm}^{-2}$ during the U Sco 2022 eruption \citep{evans2023}, 
    and $N(\mathrm{H})$ = 2--2.7 $\times\ 10^{21}\ \mathrm{cm}^{-2}$ during the U Sco 2010 eruption \citep{orio2013}.
    Therefore, the inhomogeneous absorption is considered to have little influence on our results regarding the soft X-ray stage.

\subsection{Optical plateau stage and accretion disk}
    The maximum size of the steady accretion disk can be approximately estimated as a function of mass ratio of a binary system 
    by calculating non-intersecting periodic orbits of a test particle in the restricted three-body problem \citep{Rtid}.  
    This is referred to as a tidal truncation radius, denoted as $R_{\mathrm{tid}}$ (see also the magenta line in figure \ref{figapp:roche}).
    In the case of U Sco, having a mass ratio of $q = 0.64$ (see Appendix 3), 
    the tidal truncation radius corresponds to $\alpha_{\mathrm{tid}} \sim 0.85$ (magenta line in figure \ref{fig03:ecl}), 
    where $R_{\mathrm{tid}} = \alpha_{\mathrm{tid}}  R_1^{*}$.
    Figure \ref{fig03:ecl} illustrates that when the optical plateau stage ended,
    the optical light-source radius shrank to the tidal truncation radius.  
    Therefore, it is reasonable to infer that the optical light source immediately after the optical plateau stage is the accretion disk.
    
    As for the light source expanded close to the L1 point during the optical plateau stage, 
    one could consider the possibility of the ejecta photosphere centered at the WD.
    However, figure \ref{fig03:ecl} illustrates that the soft X-ray photons were evident during the optical plateau stage.
    Judging from the evolution of the nova eruption,
    it indicates that the photosphere should have shrunk to the size of WD, smaller than the primary's Roche volume radius.
    This fact has already been shown both observationally \citep{ness2012} and theoretically (\citealp{hachisu2000}; see also Appendix 3).
    \citet{ness2012} have argued that 
    the continuum component of the SSS emission during the U Sco 2010 eruption was reproduced via achromatic Thomson scattering,
    which can increase the geometric size of the emission region without changing the shape of the continuum.
    They have also suggested that 
    the photosphere around the WD, from which the intrinsic continuum originated, 
    must have been significantly smaller than the primary's Roche lobe to avoid super Eddington luminosity.
    Therefore, it is unlikely that the optical light source expanded close to the L1 point is the photosphere. 
    
    The alternative interpretation is the expanded accretion disk.
    \citet{hachisu2000} have indeed theoretically demonstrated that the optical plateau stage during the U Sco 1999 eruption can be reproduced 
    by the combination of a slightly irradiated secondary star and a fully irradiated flaring-up accretion disk with $\alpha \sim 1.4$.
    Our observational results align quite well with this statement that the accretion disk expanded 
    and emitted strong radiation in the optical wavelengths irradiated by the hotter WD (SSS) during the optical plateau stage.

    As discussed in subsection \ref{subsec04:01},
    the optical plateau stage started later and had shorter durations in the 2022 eruption than that in the 2010 eruption.
    This follows since the optical light source during the optical plateau stage is the accretion disk, 
    and the disk can emit strong emission owing to the irradiation by the hotter WD.
    Based on the discussion in subsection \ref{subsec04:01}, 
    the recession of the photosphere might take a longer time, 
    and the irradiation of the expanded accretion disk by the hotter WD might be initiated later 
    in the 2022 eruption compared to that in the 2010 eruption.
    Also, the duration of this irradiation might be shorter 
    because of the difficulty in maintaining a high temperature of the hotter WD in the 2022 eruption.

    In the U Sco 2022 eruption, the primary eclipse was first observed on day around 11 after the optical peak, 
    followed by the beginning of the optical plateau stage on day around 14.
    This sequence of events can be explained by the photosphere receding toward the WD.
    Obviously, the secondary star emerges from the photosphere at first, followed by the emergence of the accretion disk.
    Therefore, it is consistent that 
    the primary eclipse caused by the secondary precedes the beginning of the optical plateau stage caused by the irradiated accretion disk.
    In the U Sco 2010 eruption, \citet{schaefer2011} reported that both events occurred at the same time.
    However, \citet{mason2012} have reanalyzed the light curve in \citet{schaefer2011} 
    and reported that the primary eclipse was partially detected from day $+$10.2 after the optical peak, 
    earlier than the beginning of the optical plateau stage.
    Our observational results in the 2022 eruption are in agreement with \citet{mason2012}, 
    supporting the idea that the irradiated accretion disk is mainly responsible for the optical plateau stage.

    Before the optical plateau stage, on day 11.6 after the optical peak, 
    the ingress and egress phases appear to be smaller than those observed during the optical plateau stage. 
    This might be attributed to the phase-averaged profile on day 11.6 being affected by the data points showing no variations due to the eclipses. 
    Therefore, both the ingress and egress phases were hard to measure properly, 
    so that both moved toward the middle of the primary eclipse, namely, the orbital phase 0.
    Another explanation is that the expansion of the accretion disk might have occurred already at that time,  
    but it was not sufficiently irradiated by the hotter WD to emit intense radiation in the optical region.
    Consequently, we observed only the photosphere receding toward the WD, with a radius smaller than the size of the expanded accretion disk. 

    The phase averaged profiles on days 19, 20.2, and 21.5 
    (9th, 10th, and 11th panels in figure \ref{fig03:phv}, respectively) show unexpected eclipse profiles.
    On days 19 and 20.2, hump-shaped variations above the magenta lines are presented at the orbital phase less than 0. 
    On day 21.5, a minimum of the relative magnitude can be detected at the orbital phase somewhat less than 0.
    The original light curve during this span is shown in figure \ref{fig03:lc_2022}(c) (black filled circles).
    The optical light curve reached its local minima, slightly deviating from the orbital phase 0
    (blue lines in figure \ref{fig03:lc_2022}(c)) between day 19 and day 21 after the optical peak.
    These peculiarities can be interpreted as the non-axisymmetric distribution of brightness in the expanded accretion disk.
    During the optical plateau stage, the optical light source is mainly dominated by the irradiated accretion disk compared to in quiescence.
    If the brightness distribution in the accretion disk is inhomogeneous and non-axisymmetric,
    the eclipse profile may exhibit some disturbances, or the center of light may vary, 
    resulting in a minimum relative magnitude deviating from the orbital phase 0.
    It should be noted that the MCMC method was used only to measure the ingress and egress phases of the eclipses, 
    and such inhomogeneous distribution of brightness has little influence on our results regarding the evolution of the disk radius.

\subsection{Structural changes in the accretion disk}
    In summary, our observational results indicate that 
    the accretion disk expanded beyond the tidal truncation radius and close to the L1 point during the optical plateau stage, 
    and then rapidly shrank to the tidal truncation radius within a few orbital periods.
    This phenomenon, observed for the first time, can be interpreted as a structural change 
    where the temporarily expanded accretion disk after the nova eruption returned to a steady state.

    Meanwhile, it is debatable whether the accretion disk survives or is entirely disrupted after the nova eruption.
    In the case of U Sco, based on three-dimensional hydrodynamic simulations, \citet{drake2010} demonstrated that 
    the accretion disk was entirely destroyed by the early blast wave originating from the thermonuclear explosion on the WD just after the eruption.
    However, observationally, \citet{mason2012} have reported that 
    the mass accretion from the secondary was already taking place at least 8 d after the optical peak in the 2010 eruption.
    They have stated that from day +8, the line spectrum displayed a narrow component, which varied in position and profile with time and orbital phase.
    This variation is considered to be attributed to the optically thick gas of the accretion stream, 
    moving around the primary WD but not yet in stable circular motion around it.
    Additionally, \citet{mkato2012} have argued that the disk was not gone but survived during the 2010 eruption,
    and that the numerical calculation of \citet{drake2010} cannot resolve an accretion disk which has a high density region near the equator.

    \citet{konig2022} have detected an X-ray flash in a very early stage of the nova eruption of YZ Ret.
    \citet{mkato2022} have analyzed this X-ray flash and concluded that the nova envelope was in hydrostatic balance at the X-ray flash just before the WD wind blew, 
    indicating no blast wave in the early stage of the nova eruption.
    Moreover, \citet{hachisu2022} have stated that a shock is generated outside the photosphere far beyond the binary system after the optical peak,
    denying the possibility of the disk being disrupted by the early blast wave.

    If the accretion disk is indeed not entirely disrupted, our observational results can simply be explained qualitatively as follows.
    \citet{hachisu2003} have stated that a large velocity difference between the WD wind and the accretion disk surface 
    drives a Kelvin-Helmholtz instability after the nova eruption.
    They have suggested that the internal density of the disk is much denser than that of the WD wind, 
    so only the very surface layer of the disk is dragged away like a free stream moving outward, 
    resulting in the expansion of the accretion disk.
    After the wind stops, the disk edge shrinks toward the tidal truncation radius 
    in several dynamical timescales, namely, several orbital periods, 
    and finally the accretion disk settles into a steady state.
    The optical plateau stage is when the WD wind gradually weakens and stops at the end as the soft X-ray photons become noticeable,
    so our observations align quite well with this model of \citet{hachisu2003}, 
    supporting the idea the accretion disk survives the eruption and expands until the WD wind stops and the optical plateau stage ends.

    Another possibility would be a change in the mass-transfer rate.
    In the case of VY Scl stars, \citet{murray2000} have suggested that 
    the suddenly reduced mass flux from the secondary star might lead to an expansion of the accretion disk, 
    after which the disk returns to its initial size on its viscous time-scale.
    They have argued that mixing with the mass-transfer stream reduces specific angular momentum in the outer disk and restricts the accretion disk radially, 
    so the lower mass-transfer rate relaxes the radial restriction on the accretion disc. 
    As for U Sco, the mass-transfer rate might be reduced suddenly after the nova eruption,
    and the accretion disk, which is fully ionized and has high viscosity, might expand close to the L1 point.  
    However, irradiation of the secondary just after the nova eruption may lead to a temporarily increased mass-transfer rate (\citealp{mroz2016}; \citealp{hillman2020}).
    If the mass-transfer rate is enhanced rather than reduced, the above argument will not hold.
    
    Although further observations and numerical simulations are indeed necessary to fully understand the behavior of the accretion disk after the nova eruption,
    our observational results may support the idea that the accretion disk is not entirely blown off.

\section{Summary}\label{sec05:sum}
    We obtained rich optical and soft X-ray photometric data in the 2022 eruption of U Sco.
    We demonstrate that the optical plateau and soft X-ray stages started later and had shorter durations in the 2022 eruption than those in the 2010 eruption.
    Moreover, the soft X-ray light curve in the 2022 eruption peaked earlier and was less luminous compared to that in the 2010 eruption.
    These observational results might be attributed to the differences in the envelope mass, especially in the hydrogen content of the envelope.
    The appearance of the primary eclipse preceded the beginning of the optical plateau stage during the 2022 eruption, 
    which we interpret as the shrinking ejecta photosphere toward the WD surface. 
    Additionally, by determining the ingress and egress timings of the eclipses, 
    we show that the optical plateau stage is well explained by the accretion disk.
    Although the accretion disk expanded close to the L1 point during the optical plateau stage,  
    it then rapidly shrank to the tidal truncation radius within just a few orbital periods after the optical plateau stage ended.
    The expanded accretion disk can be explained by the Kelvin-Helmholtz instability 
    caused by a large velocity difference between the WD wind and the accretion disk surface,
    suggesting that the accretion disk may not be entirely blown off even right after the nova eruption.
    
\section*{Supplementary data} 
    The following supplementary data is available in the online version of this article. 
    
    E-tables 1--4.

\begin{ack}

    This work was supported by many VSNET and VSOLJ observers. 
    We thank world-wide professional and amateur observers who have shared data with VSNET and VSOLJ.
    This work was financially supported by the Japan Society for the Promotion of Science Grants-in-Aid for Scientific Research (KAKENHI) Grant Numbers 21J22351 (Y.T.), 21K03616 (D.N. and T.K.).
    This work was supported by Slovak Research and Development Agency under contract No. APVV-20-0148.
    We are thankful to I. Hachisu for providing us with valuable comments regarding the behavior of an accretion disk and blast wave after the nova eruption.
    We acknowledge with thanks the variable star observations from the AAVSO International Database contributed by observers worldwide and used in this research.
    Based on observations obtained with the Samuel Oschin 48-inch Telescope at the Palomar Observatory as part of the Zwicky Transient Facility project. ZTF is supported by the National Science Foundation under Grant No. AST-1440341 and a collaboration including Caltech, IPAC, the Weizmann Institute for Science, the Oskar Klein Center at Stockholm University, the University of Maryland, the University of Washington, Deutsches Elektronen-Synchrotron and Humboldt University, Los Alamos National Laboratories, the TANGO Consortium of Taiwan, the University of Wisconsin at Milwaukee, and Lawrence Berkeley National Laboratories. Operations are conducted by COO, IPAC, and UW.
    ASAS-SN is supported by the Gordon and Betty Moore Foundation through grant GBMF5490 to the Ohio State University and NSF grant AST-1515927. Development of ASAS-SN has been supported by NSF grant AST-0908816, the Mt. Cuba Astronomical Foundation, the Center for Cosmology and AstroParticle Physics at the Ohio State University, the Chinese Academy of Sciences South America Center for Astronomy (CASSACA), the Villum Foundation, and George Skestos.
    This work has made use of data from the Asteroid Terrestrial-impact Last Alert System (ATLAS) project. The Asteroid Terrestrial-impact Last Alert System (ATLAS) project is primarily funded to search for near earth asteroids through NASA grants NN12AR55G, 80NSSC18K0284, and 80NSSC18K1575; byproducts of the NEO search include images and catalogs from the survey area. This work was partially funded by Kepler/K2 grant J1944/80NSSC19K0112 and HST GO-15889, and STFC grants ST/T000198/1 and ST/S006109/1. The ATLAS science products have been made possible through the contributions of the University of Hawaii Institute for Astronomy, the Queen’s University Belfast, the Space Telescope Science Institute, the South African Astronomical Observatory, and The Millennium Institute of Astrophysics (MAS), Chile.

\end{ack}

\begin{appendices}
\def\thesection{Appendix \arabic{section}.}
\section{Phase-averaged profile}
    To demonstrate when the optical eclipses emerged, we created phase-averaged optical light curves of the U Sco 2022 eruption as below.
    
    First of all, we extracted the localized luminosity variations due to the eclipses from the overall light curve.
    The global trend was calculated by locally weighted polynomial regression
    \citep[LOWESS;][]{LOWESS} for the out-of-eclipse data points during the eruption.
    We subtracted this trend from the observed light curve to obtain relative magnitudes.

    Next, we sampled the data every two epochs, centered at the orbital phase 0.5 after the optical peak,
    and plotted the orbital phase on the x-axis and the relative magnitude on the y-axis as follows, 
    representing the luminosity variations caused by the eclipses.
    The orbital period of U Sco is derived in \citet{schaefer2011}, as described in section \ref{sec01:intro},
    and we assumed that it remained constant during the U Sco 2010 and 2022 eruptions.
    \citet{schaefer2022} reported a slight increase in the orbital period during the 2022 eruption compared to that during the 2010 eruption.
    However, this increase in the orbital period did not result in any significant differences in our results.
    Epochs of the middle of the primary eclipse are also derived in \citet{schaefer2011}, as described in equation (\ref{epoch}), 
    and we adopted these epochs during the U Sco 2022 eruption to obtain orbital phases.
    The sampling data were binned to 1$/$60 phases, and we calculated the average of relative magnitudes for each phase, creating a phase-averaged profile.

\section{MCMC analysis}
    \begin{figure}[tb]
        \begin{center}
            \includegraphics[width=\linewidth]{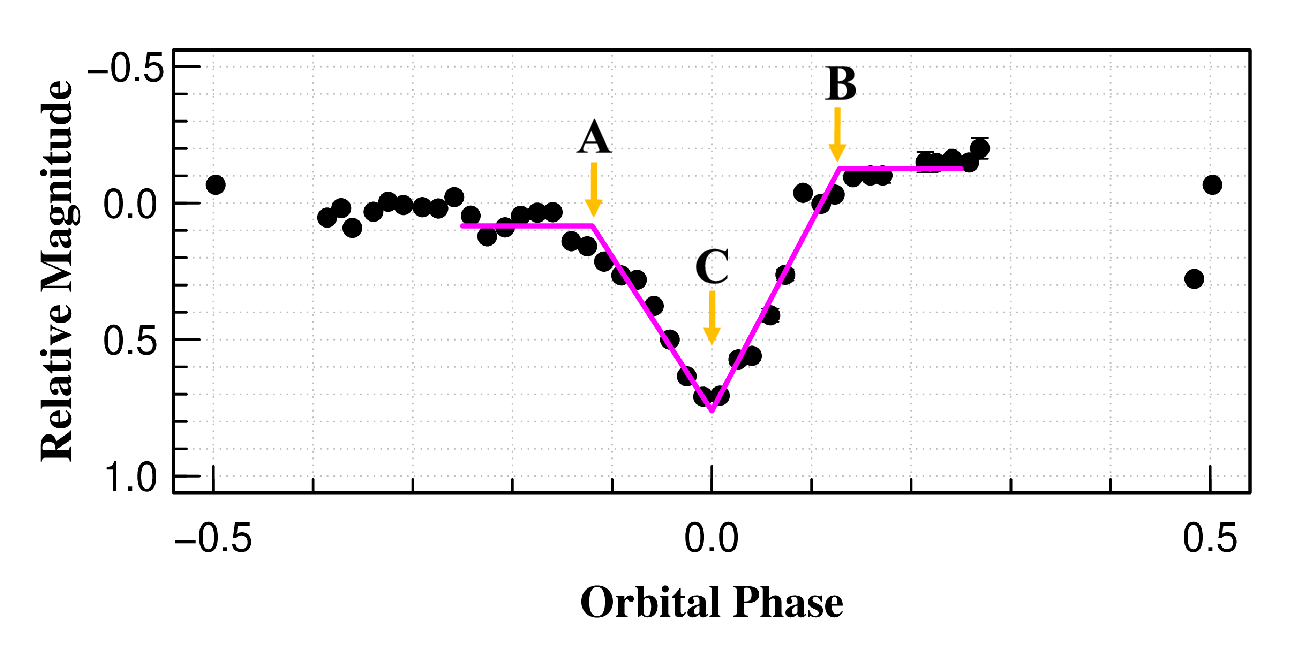}
        \end{center}
        \caption{An example of parameter estimation using the MCMC method 
        (for a phase-averaged profile centered on day 26.4 after the optical peak during the U Sco 2022 eruption).
        The magenta line represent the regression model with the determined parameters, 
        bending at the points $A(a, m_a)$, $B(b, m_b)$, and $C(0, m_0)$.}\label{figapp:mcmc}
    \end{figure}
    
    For each phase-averaged profile, we determined the ingress and egress phases of the primary eclipse using the Markov Chain Monte Carlo (MCMC) method.
    The parameter space is $\theta = \{a, m_a, b, m_b, m_0\}$.
    The parameters $(a, m_a)$ represent the eclipse ingress phase and its relative magnitude.
    The parameters $(b, m_b)$ represent the eclipse egress phase and its relative magnitude.
    The parameter $m_0$ represents the relative magnitude at the middle of the eclipse, namely, the orbital phase 0.
    For a orbital phase $x_i$ which satisfies $-0.25 \leq x_i \leq 0.25$, 
    the actual observed relative magnitude is denoted as $y_{\mathrm{obs}}(x_i)$, and the regression model is computed as follows:

    \begin{equation}
        y_{\mathrm{model}}(x_i) = \left\{
        \begin{array}{lll}
            m_a & \mathrm{if} & -0.25 \leq x_i < a, \\
            m_a - \frac{(m_0 - m_a)}{a}(x_i-a) & \mathrm{if} & a \leq x_i < 0, \\
            m_b - \frac{(m_0 - m_b)}{b}(x_i-b) & \mathrm{if} & 0 \leq x_i < b, \\
            m_b & \mathrm{if} & b \leq x_i \leq 0.25.
        \end{array}
        \right.
    \end{equation}
    The likelihood function can be written as
    \begin{equation}
        \mathcal{L}(\theta) = \prod_i \frac{1}{\sqrt{2\pi\sigma^2}}\exp\left\{-\frac{[y_{\mathrm{obs}}(x_i) - y_{\mathrm{model}}(x_i)]^2}{2\sigma^2}\right\},
    \end{equation}
    where the parameter $\sigma$ is assumed to be $\sigma = 0.01$.
    We especially determined the parameters $(a, b)$, the ingress and egress phases of the eclipses.
    Figure \ref{figapp:mcmc} shows an example.

\section{Light-source radius}
    \begin{figure}[tb]
        \begin{center}
            \includegraphics[width=\linewidth]{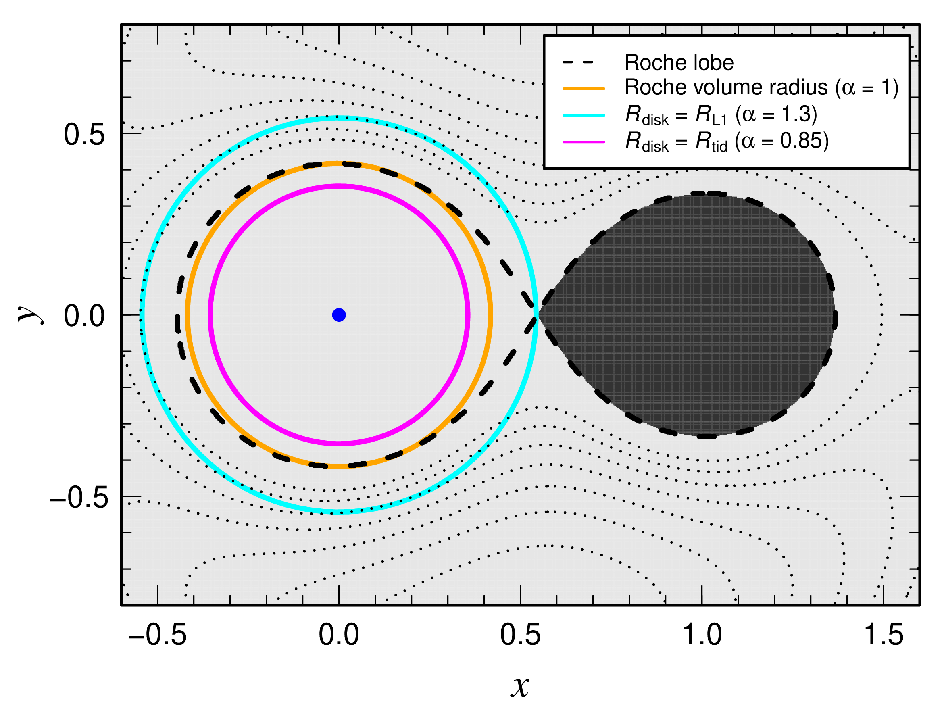}
        \end{center}
        \caption{Schematic illustration of the U Sco binary system in its orbital plane normalized by the binary separation.
        The blue filled circle centered at $(0,0)$ represents the primary WD and the ejecta photosphere.
        The black region represents the secondary star filling its Roche lobe (black dashed line).
        The axisymmetric accretion disk centered at the WD expands to have an outer radius equal to the primary's Roche volume radius $R_1^{*}$ (orange line), 
        the distance between the L1 point and the primary WD $R_{\mathrm{L1}}$ (cyan line), 
        and the tidal truncation radius $R_{\mathrm{tid}}$ (magenta line).
        Black dotted lines represent the Roche equipotentials.}\label{figapp:roche}
    \end{figure}
    
    At the ingress and egress phases of the primary eclipse, 
    the secondary star just touches the eclipsed optical light source geometrically, projected onto the plane of the sky.
    Assuming a geometric configuration of the binary system for a given light-source radius, 
    there exist some line-of-sight directions, namely, orbital phases in which the secondary and the light source just intersect on the projected plane. 
    Therefore, we estimated the outer radius of the optical light source centered at the primary WD in reverse, 
    based on the ingress and egress phases determined by the MCMC method.

    We assumed a binary parameter of U Sco as below 
    (see also figure \ref{figapp:roche} for a schematic illustration of the U Sco binary system in its orbital plane).
    The primary WD mass is $M_1 = 1.37\ \MO$, the secondary mass is $M_2 = 0.88\ \MO$, the orbital inclination is $82.7\degree$, and the orbital period is $1.23$ d,
    as described in section \ref{sec01:intro}.
    The secondary star is assumed to fill its Roche lobe (black region in figure \ref{figapp:roche}).
    The WD and secondary are synchronously rotating around the common center of mass on a circular orbit with a binary separation $a = 6.33\ \RO$, 
    as determined from the orbital period.
    \citet{hachisu2000} have theoretically demonstrated that 
    the ejecta photosphere centered at the WD shrank to $R_{\mathrm{ph}} \sim 0.1\ \RO$ on day 14 after the optical peak,
    and drastically shrank to $R_\mathrm{ph} \sim 0.003\ \RO$ on day 18, when the WD wind stopped during the U Sco 1999 eruption.
    Here, the photosphere is assumed to be a spherical body centered at the WD and have a radius $0.1\ \RO$ (blue filled circle in figure \ref{figapp:roche}), 
    much smaller than the binary size ($R_{\mathrm{ph}} \sim 0.04 R_1^{*}$, based on the definition below).
    
    As for the accretion disk, we also adopted the model results of \citet{hachisu2000} as follows.
    The accretion disk is axisymmetric, centered at the primary WD, and has an outer radius given by

    \begin{equation}
        R_{\mathrm{disk}} = \alpha R_1^{*}.
    \end{equation}
    The $R_1^{*}$ is the primary's Roche volume radius (orange line in figure \ref{figapp:roche}),
    and a sphere with this radius has the same volume as the primary's Roche lobe.
    It can be approximately estimated that
    
    \begin{equation}
        R_1^{*}(q) = \frac{0.49 q^{-2/3}}{0.6 q^{-2/3} + \ln(1+q^{-1/3})}a,
    \end{equation}
    where $q$ is the mass ratio of a binary system, defined by $q\equiv M_1/M_2$, and $a$ is the binary separation \citep[e.g.,][]{eggleton1983}. 
    In the case of U Sco, it is calculated to be $R_1^{*} = 2.64 R_{\solar}$ with a mass ratio of $q = 0.64$.
    In addition, the disk has a thickness given by
    \begin{equation}
        h = \beta R_{\mathrm{disk}}\left(\frac{\varpi}{R_\mathrm{disk}}\right)^2,
    \end{equation}
    where $h$ is the height of the surface from the equatorial plane, and $\varpi$ is the distance on the equatorial plane from the center of the WD.
    \citet{hachisu2000} have theoretically demonstrated that $\beta = 0.30$ when the WD wind blew, and $\beta = 0.35$ after it stopped.
    
    For the ingress and egress phases determined for each phased-averaged profile, 
    we estimated $\alpha$ as a light-source radius, fixing $\beta = 0.30$.
    We tested a model of thinner accretion disk, $\beta \sim 0.10$ (i.e., standard disk model), but this change did not affect our results.
    This is because, the inclination of U Sco is so high that 
    the eclipse ingress and egress times are considered to be primarily determined by the size of the accretion disk in its orbital plane, but not the vertical extension.
    We also investigated other configurations, such as the disk being an infinitely long cylinder,
    or the spherical photosphere expanding beyond the disk (i.e., $R_\mathrm{disk} =  0.1\ \RO$ and $R_\mathrm{ph} = \alpha R_1^{*}$).
    However, these configurations did not result in any significant differences in our results.
    \citet{schaefer2011} also estimated the optical emitting region of the U Sco 2010 eruption in a similar way.
    However, it should be noted that they considered a spherical secondary star with a radius of the secondary's Roche volume radius 
    and might overestimate the light-source radius compared to our own model.
    
\end{appendices}

\bibliographystyle{pasjlike}
\bibliography{cvs}

\newcommand{\noop}[1]{}
\begin{thebibliography}{}

\bibitem[Bellm  et~al.(2019)]{ZTF}
  Bellm, E.~C., {et~al.}\ 2019, PASP, 131, 018002

\bibitem[Brandi et~al.(2009)]{brandi2009}
  Brandi, E., Quiroga, C., Mikołajewska, J., Ferrer, O.~E., \& García, L.~G.\
  2009, A\&A, 497, 815

\bibitem[Burrows  et~al.(2005)]{XRT}
  Burrows, D.~N., {et~al.}\ 2005, Space\ Sci.\ Rev., 120, 165

\bibitem[Chomiuk et~al.(2021)]{chomiuk2021}
  Chomiuk, L., Metzger, B.~D., \& Shen, K.~J.\ 2021, Annual Review of Astronomy
  and Astrophysics, 59, 391

\bibitem[Cleveland(1979)]{LOWESS}
  Cleveland, W.~S.\ 1979, Journal of the American Statistical Association, 74,
  829

\bibitem[Drake and Orlando(2010)]{drake2010}
  Drake, J.~J., \& Orlando, S.\ 2010, ApJ, 720, L195

\bibitem[Eggleton(1983)]{eggleton1983}
  Eggleton, P.~P.\ 1983, ApJ, 268, 368

\bibitem[Evans et~al.(2023)]{evans2023}
  Evans, A., Banerjee, D. P.~K., Woodward, C.~E., Geballe, T.~R., Gehrz, R.~D.,
  Page, K.~L., \& Starrfield, S.\ 2023, MNRAS, 522, 4841

\bibitem[Frank et~al.(2002)]{frank2002}
  Frank, J., King, A., \& Raine, D.\ 2002, Accretion Power in Astrophsics
  (Cambridge: Cambridge University Press)

\bibitem[Fujimoto and Iben(1992)]{fujimoto1992}
  Fujimoto, M.~Y., \& Iben, I.\ 1992, ApJ, 399, 646

\bibitem[Gallagher and Starrfield(1978)]{Nova}
  Gallagher, J.~S., \& Starrfield, S.\ 1978, Annual Review of Astronomy and
  Astrophysics, 16, 171

\bibitem[Gehrels  et~al.(2004)]{Swift}
  Gehrels, N., {et~al.}\ 2004, ApJ, 611, 1005

\bibitem[Hachisu and Kato(2003)]{hachisu2003}
  Hachisu, I., \& Kato, M.\ 2003, ApJ, 588, 1003

\bibitem[Hachisu and Kato(2006a)]{hachisu2006}
  Hachisu, I., \& Kato, M.\ 2006a, ApJ, 642, L53

\bibitem[Hachisu and Kato(2006b)]{hachisu2006_2}
  Hachisu, I., \& Kato, M.\ 2006b, ApJS, 167, 59

\bibitem[Hachisu and Kato(2022)]{hachisu2022}
  Hachisu, I., \& Kato, M.\ 2022, ApJ, 939, 1

\bibitem[Hachisu et~al.(2000)]{hachisu2000}
  Hachisu, I., Kato, M., Kato, T., \& Matsumoto, K.\ 2000, ApJ, 528, L97

\bibitem[Hachisu et~al.(2007)]{hachisu2007}
  Hachisu, I., Kato, M., \& Luna, G. J.~M.\ 2007, ApJ, 659, L153

\bibitem[Henze  et~al.(2018)]{henze2018}
  Henze, M., {et~al.}\ 2018, ApJ, 857, 68

\bibitem[Hillman et~al.(2020)]{hillman2020}
  Hillman, Y., Shara, M.~M., Prialnik, D., \& Kovetz, A.\ 2020, Nature Astron.,
  4, 886

\bibitem[Kato and Hachisu(2012)]{mkato2012}
  Kato, M., \& Hachisu, I.\ 2012, Bull.\ Astron.\ Soc.\ India, 40, 393

\bibitem[Kato and Hachisu(2020)]{mkato2020}
  Kato, M., \& Hachisu, I.\ 2020, PASJ, 72, 82

\bibitem[Kato et~al.(2022)]{mkato2022}
  Kato, M., Saio, H., \& Hachisu, I.\ 2022, ApJLett, 935, L15

\bibitem[Kato et~al.(2004)]{VSNET}
  Kato, T., Uemura, M., Ishioka, R., Nogami, D., Kunjaya, C., Baba, H., \&
  Yamaoka, H.\ 2004, PASJ, 56, S1

\bibitem[Kochanek  et~al.(2017)]{ASN_2}
  Kochanek, C.~S., {et~al.}\ 2017, PASP, 129, 104502

\bibitem[König  et~al.(2022)]{konig2022}
  König, O., {et~al.}\ 2022, Nature, 605, 248

\bibitem[Leahy and Leahy(2015)]{leahy2015}
  Leahy, D.~A., \& Leahy, J.~C.\ 2015, Computational Astrophysics and
  Cosmology, 2, 4

\bibitem[Mason et~al.(2012)]{mason2012}
  Mason, E., Ederoclite, A., Williams, R.~E., Valle, M.~D., \& Setiawan, J.\
  2012, A\&A, 544, A149

\bibitem[Maxwell et~al.(2014)]{maxwell2014}
  Maxwell, M.~P., Rushton, M.~T., \& Eyres, S. P.~S.\ 2014, ASP\ Conf.\ Ser.\,
  490, 205

\bibitem[Mróz  et~al.(2016)]{mroz2016}
  Mróz, P., {et~al.}\ 2016, New\ Astron., 537, 649

\bibitem[Murray et~al.(2000)]{murray2000}
  Murray, J.~R., Warner, B., \& Wickramasinghe, D.~T.\ 2000, MNRAS, 315, 707

\bibitem[Ness  et~al.(2012)]{ness2012}
  Ness, J.~U., {et~al.}\ 2012, ApJ, 745, 43

\bibitem[Ness  et~al.(2023)]{ness2023}
  Ness, J.~U., {et~al.}\ 2023, A\&A, 670, A131

\bibitem[Orio  et~al.(2013)]{orio2013}
  Orio, M., {et~al.}\ 2013, MNRAS, 429, 1342

\bibitem[Paczynski(1977)]{Rtid}
  Paczynski, B.\ 1977, ApJ, 216, 822

\bibitem[Page  et~al.(2022a)]{page2022_3}
  Page, K.~L., {et~al.}\ 2022a, MNRAS, 514, 1557

\bibitem[Page et~al.(2022b)]{page2022}
  Page, K.~L., Evans, P.~A., \& Beardmore, A.~P.\ 2022b, Astron.\ Telegram,
  15438, 1

\bibitem[Page and Osborne(2022)]{page2022_2}
  Page, K.~L., \& Osborne, J.~P.\ 2022, Astron.\ Telegram, 15465, 1

\bibitem[Schaefer(2010)]{schaefer2010_1}
  Schaefer, B.~E.\ 2010, ApJ, 373, 187

\bibitem[Schaefer(2022)]{schaefer2022}
  Schaefer, B.~E.\ 2022, MNRAS, 516, 4497

\bibitem[Schaefer  et~al.(2010a)]{schaefer2010}
  Schaefer, B.~E., {et~al.}\ 2010a, AJ, 140, 925

\bibitem[Schaefer  et~al.(2010b)]{schaefer2010_2}
  Schaefer, B.~E., {et~al.}\ 2010b, Astron.\ Telegram, 2452, 1

\bibitem[Schaefer  et~al.(2011)]{schaefer2011}
  Schaefer, B.~E., {et~al.}\ 2011, ApJ, 742, 113

\bibitem[Schlegel  et~al.(2010a)]{schlegel2010}
  Schlegel, E.~M., {et~al.}\ 2010a, Astron.\ Telegram, 2419, 1

\bibitem[Schlegel  et~al.(2010b)]{schlegel2010_2}
  Schlegel, E.~M., {et~al.}\ 2010b, Astron.\ Telegram, 2430, 1

\bibitem[Shappee  et~al.(2014)]{ASN_1}
  Shappee, B.~J., {et~al.}\ 2014, ApJ, 788, 48

\bibitem[Starrfield et~al.(1972)]{TNR}
  Starrfield, S., Truran, J.~W., Sparks, W.~M., \& Kutter, G.~S.\ 1972, ApJ,
  176, 169

\bibitem[Thoroughgood et~al.(2001)]{thoroughgood2001}
  Thoroughgood, T.~D., Dhillon, V.~S., Littlefair, S.~P., Marsh, T.~R., \&
  Smith, D.~A.\ 2001, MNRAS, 327, 1323

\bibitem[Tonry  et~al.(2018)]{ATLAS}
  Tonry, J.~L., {et~al.}\ 2018, PASP, 130, 064505

\bibitem[Warner(1995)]{warner}
  Warner, B.\ 1995, Cataclysmic Variable Stars (Cambridge: Cambridge University
  Press)

\bibitem[Wolf et~al.(2013)]{wolf2013}
  Wolf, W.~M., Bildsten, L., Brooks, J., \& Paxton, B.\ 2013, ApJ, 777, 136

\end{thebibliography}

\end{document}